\documentclass{emulateapj}
\usepackage{graphicx}
\usepackage{epstopdf}
\usepackage{epsfig}

\newcommand{\avg}[1]{\ensuremath{\langle #1 \rangle}}
\newcommand{\bma}{\begin{math}}
\newcommand{\ema}{\end{math}}
\newcommand{\beq}{\begin{equation}}
\newcommand{\eeq}{\end{equation}}
\newcommand{\beqa}{\begin{eqnarray}}
\newcommand{\eeqa}{\end{eqnarray}}
\newcommand{\bc}{\begin{center}}
\newcommand{\ec}{\end{center}} 
\newcommand{\bit}{\begin{itemize}}
\newcommand{\eit}{\end{itemize}}

\font\BFd=cmmib10
\font\BFt=cmmib10
\font\BFs=cmmib10 scaled 700
\font\BFss=cmmib10 scaled 500

\def\bbox#1{%
\relax\ifmmode
\mathchoice
{{\hbox{\BFd #1}}}
{{\hbox{\BFt #1}}}
{{\hbox{\BFs #1}}}
{{\hbox{\BFss #1}}}
\else \mbox{#1} \fi }

\def\k{{\bbox{k}}}
\def\rvec{{\bbox{r}}}

\def\x{{\bbox{x}}}

\begin{document}



 
\submitted{\today. To be submitted to \apj.} 

\title{Probing Reionization with the 21 cm-Galaxy Cross Power Spectrum}
\author{Adam Lidz\altaffilmark{1}, Oliver Zahn\altaffilmark{2}, Steven R. Furlanetto\altaffilmark{3}, Matthew McQuinn\altaffilmark{1},
Lars Hernquist\altaffilmark{1}, Matias Zaldarriaga\altaffilmark{1,4}}
\altaffiltext{1}{Harvard-Smithsonian Center for Astrophysics, 60 Garden Street,
Cambridge, MA 02138, USA}
\altaffiltext{2}{Berkeley Center for Cosmological Physics, Department of Physics,
University of California, and Lawrence Berkeley National Labs, 1 Cyclotron Road, Berkeley,
CA 94720, USA}
\altaffiltext{3}{Department of Physics and Astronomy, University of California, Los Angeles,
CA 90095, USA}
\altaffiltext{4}{Jefferson Laboratory of Physics; Harvard University; Cambridge, MA 02138, USA}
\email{alidz@cfa.harvard.edu}

\begin{abstract}
The cross-correlation between high redshift galaxies and 21 cm emission
from the high redshift intergalactic medium (IGM) promises to be an 
excellent probe of the Epoch of Reionization (EoR). 
On large scales, the 21 cm
and galaxy fields are anti-correlated during most of the reionization
epoch.  However, 
on scales smaller than the size of the H II regions around detectable galaxies, the 
two fields become roughly uncorrelated. 
Consequently, 
the 21 cm-galaxy cross power spectrum provides a tracer of bubble
growth during reionization, with the signal turning over on progressively
larger scales as reionization proceeds. The precise turnover scale depends on the
minimum host mass of the detectable galaxies, and the galaxy selection technique. Measuring 
the turnover scale as a function of galaxy luminosity constrains the
characteristic bubble size around galaxies of different luminosities.
The cross spectrum becomes positive on small scales if ionizing photons fail to escape
from low mass galaxies, and these galaxies are detectable longward of the hydrogen 
ionization edge, because in this case some 
identifiable galaxies lie outside of ionized regions.
LOFAR can potentially measure the 21 cm-galaxy
cross spectrum in conjunction with mild extensions to the existing Subaru survey for
$z=6.6$ Lyman-alpha emitters, while the
MWA is slightly less sensitive for detecting the cross spectrum. A futuristic galaxy
survey covering a sizable fraction of the MWA field of view ($\sim 800$ deg$^2$)
can probe the scale dependence of the cross spectrum, constraining the filling factor of H II
regions at different redshifts during reionization, and providing other valuable constraints
on reionization models. 
 \end{abstract}


\section{Introduction} \label{sec:intro}

Detecting 21 cm emission from the high redshift intergalactic medium (IGM)
will potentially revolutionize our understanding of cosmic reionization.
The 21 cm signal promises direct, three-dimensional information regarding the
state of the IGM during reionization (e.g. Scott \& Rees 1990, Madau et al. 1997, 
Furlanetto et al. 2006a).
Unfortunately, experimental challenges are substantial. In particular, astrophysical
foregrounds are expected to be four to five orders of magnitude larger
than the signal from the high redshift IGM. However, {\it known} foreground 
contaminants are spectrally smooth and should be distinguishable from the high redshift
21 cm signal itself (Zaldarriaga et al. 2004). 

Given that we anticipate observational complications, it is important to
develop diagnostics to confirm that the detected 21 cm signal indeed originates from
the high redshift IGM. 
One such approach is to measure the cross correlation between
21 cm emission and a high redshift galaxy survey 
(Furlanetto \& Lidz 2007, Wyithe \& Loeb 2007).
Since most of the anticipated foregrounds come from low redshift -- primarily 
galactic synchrotron -- and not
from high redshift galaxies, the mean 21 cm-galaxy cross power spectrum signal is largely
immune to foreground contamination (Furlanetto \& Lidz 2007). Detecting a 
21 cm-galaxy cross correlation
should hence confirm that the detected 21 cm signal comes from 
the high redshift IGM. Moreover, continuing efforts
are pushing galaxy surveys towards higher redshifts, and it is natural to consider the 
information
that may be gleaned from combining galaxy and 21 cm surveys. Detecting galaxies at very high
redshift is extremely challenging (e.g. Stark et al. 2007, Bouwens et al. 2008), but
we will show here that a cross spectrum detection may already be possible with 
modest extensions 
to the Subaru survey (Kashikawa et al. 2006; see 
also Wyithe \& Loeb 2007, Furlanetto \& Lidz 2007).

In addition to these important practical advantages, the 21 cm-galaxy cross 
correlation is potentially sensitive to the size and filling factor of H II regions,
the clumpiness of the IGM, and the nature of the ionizing sources. 
The 21 cm-galaxy cross correlation will also provide a more direct tracer 
of the interplay
between the reionizing sources and the surrounding IGM, than the 21 cm auto power 
spectrum. The 
21 cm-galaxy cross correlation should hence provide a unique and powerful
probe of the Epoch of Reionization (EoR) and early structure formation. Here we follow up on earlier work
by Wyithe \& Loeb (2007) and Furlanetto \& Lidz (2007), and 
focus on modeling the 21 cm-galaxy cross power spectrum, and exploring the 
insights that future surveys will provide regarding the EoR.

The outline of this paper is as follows. In \S \ref{sec:cross} we establish
notation, describe the reionization simulations used in our analysis, and
examine the basic simulated signal.
We then characterize (\S \ref{sec:evol_sig}) the dependence of the cross spectrum 
on redshift and
ionization fraction. In \S \ref{sec:sources} we illustrate how the 
signal is sensitive to the properties of the ionizing sources. In \S \ref{sec:recombs} we
describe its variation with the abundance of Lyman-limit systems. We then examine
the signal's dependence on the way in which high redshift galaxies are selected 
(\S \ref{sec:selection}), contrasting the results for Ly-$\alpha$ selected galaxies
with galaxies selected through other techniques. 
In \S \ref{sec:detectability} we briefly discuss the statistical power of
future surveys to constrain reionization through measurements of the
21 cm-galaxy cross power spectrum. Finally, we summarize our main
results and conclude in \S \ref{sec:conclusions}.

Throughout we consider a $\Lambda$CDM cosmology parameterized by:  
$n_s = 1$, $\sigma_8 = 0.8$,
$\Omega_m = 0.27$, $\Omega_\Lambda = 0.73$, $\Omega_b = 0.046$, and $h=0.7$, 
(all symbols
have their usual meanings), consistent with the WMAP constraints from 
Spergel et al. (2007) and Komatsu et al. (2008).

\section{The 21 cm-galaxy cross power spectrum}
\label{sec:cross}

In this paper we focus on the cross power spectrum between the 21 cm and galactic abundance
fields. Each field is non-Gaussian and so the cross and auto power spectra 
alone provide 
an incomplete description of the fields' statistical properties. 
However, the limited sensitivity of
first generation 21 cm surveys will prohibit detailed imaging of the 21 cm 
field (McQuinn et al. 2006), and we expect these surveys to have relatively low signal
to noise for detecting higher order moments of the 21 cm field. We hence focus on the
cross power spectrum throughout. 

In order to explore the information content of the 21 cm-galaxy cross power spectrum, it
is useful to decompose the signal into the sum of several contributing terms.
Throughout this work we adopt the limit that the spin temperature of 
the 21 cm transition is much higher than the
CMB temperature globally (Ciardi \& Madau 2003, Pritchard \& Furlanetto 2007), $T_s >> T_{\rm CMB}$, and we ignore peculiar velocities -- which should be a good approximation
during most of the reionization epoch (Mesinger \& Furlanetto 2007a).
With these assumptions the 21 cm-galaxy cross power spectrum can be written as:
\beqa
\Delta^2_{\rm 21, gal}(k) = && \tilde{\Delta}^2_{\rm 21, gal}(k)/T_0 = \avg{x_{\rm HI}}
[\Delta^2_{\rm x, gal}(k) \nonumber \\ && + \Delta^2_{\rm \rho, gal}(k) 
+ \Delta^2_{\rm x \rho, gal}(k)].
\label{eq:p21_gal_cross}
\eeqa
In this equation $\Delta^2_{\rm 21, gal}(k)$ denotes the cross power 
spectrum
between the 21 cm field and the galaxy overdensity at wavenumber
$k=|\k|$. The 21 cm field at spatial
position $\rvec$
is given by $\delta_T (\rvec) = T_0 \avg{x_H} \left(1 + \delta_x(\rvec)\right) 
\left(1 + \delta_\rho(\rvec)\right)$;
$T_0$ is the 21 cm brightness temperature, relative to the CMB, at
the redshift in question for a fully neutral gas element at the cosmic
mean density, and $\avg{x_H}$ is the volume-averaged hydrogenic
neutral fraction. The field $\delta_x(\rvec) = (x_H(\rvec) - \avg{x_H})/\avg{x_H}$
is the fractional fluctuation in the neutral hydrogen fraction, while
$\delta_\rho$ is the fractional gas density fluctuation.
Similarly $\delta_g(\rvec) = (n_g(\rvec) - \avg{n_g})/\avg{n_g}$ is the fractional fluctuation in galaxy abundance, where $n_g(\rvec)$ 
specifies the co-moving number density of galaxies at spatial position
$\rvec$, and $\avg{n_g}$ denotes
the volume-averaged galactic abundance. Our notation labels the dimensionless cross 
spectrum of two random fields, $a$ and $b$, by $\Delta^2_{\rm a, b}(k) =
k^3 P_{\rm a, b}(k)/(2 \pi^2)$
and $\Delta^2_{\rm x, gal}$, for example, is shorthand for the cross power
spectrum between $\delta_x$ and $\delta_g$ (and $P_{\rm a, b}$ is the usual dimensionful
cross spectrum).
We use a similar shorthand for $\Delta^2_{\rm \rho, gal}(k)$ and
$\Delta^2_{\rm x \rho, gal}(k)$. Throughout we work with the power spectrum of the dimensionless
field $\delta_T(\rvec)/T_0$ which we denote by $\Delta^2_{\rm 21, gal}(k)$, which is distinguished
from the dimensionful power spectrum $\tilde{\Delta}^2_{\rm 21, gal}(k)$ by the factor of $T_0$
as in Equation \ref{eq:p21_gal_cross}.

The individual terms contributing to the 21 cm-galaxy cross power spectrum have the 
following 
physical interpretations. The first term, $\Delta^2_{\rm x, gal}(k)$, represents the
cross power spectrum between the neutral hydrogen fraction and galaxy density fields. The second 
term,
$\Delta^2_{\rm \rho, gal}(k)$, is the cross power spectrum between the matter
and galaxy overdensity fields. The final term, $\Delta^2_{\rm x \rho, gal}(k)$,
is a three field term which would vanish if each contributing field were Gaussian. In fact, 
we show below that this term is generally significant during reionization (see Lidz et al. 2007a
for discussion of a similar term, $\Delta^2_{\rm x \rho, \rho}(k)$,
which contributes to the 21 cm auto power spectrum). 
We will sometimes refer to these terms respectively as 
the {\em `x-gal'}, {\em `$\rho-gal$'}, and {\em `three-field'} terms.
Let us examine each contributing term
from our reionization simulations.

\subsection{Reionization Simulations}
\label{sec:sims}

First, we briefly describe the two types of reionization simulations used in this work. The first type 
are the reionization simulations of McQuinn et al. (2007b). 
In these simulations, radiative transfer
is treated in a post-processing stage using the code of McQuinn et al. (2007a), a refinement of
the Sokasian et al. (2001, 2003) code,
which in turn uses the adaptive ray-tracing scheme of Abel \& Wandelt (2002).
The radiative transfer calculation is performed on top of a $130$ Mpc/$h$, $1024^3$ particle dark
matter simulation run with an enhanced version of Gadget-2 (Springel 2005). The minimum resolved
halo in this simulation is $\sim 10^{10} M_\odot$, but smaller mass halos down to the atomic 
cooling mass (Barkana \& Loeb 2001), $M_{\rm cool} \sim 10^8 M_\odot$, are incorporated with
the appropriate abundance and clustering as in McQuinn et al. (2007a). Ionizing sources 
are placed
in simulated halos with simple prescriptions. In our fiducial model, we assume that a source's 
ionizing luminosity is proportional to its host halo mass. We assume that gas directly
traces the dark matter, which should be a good approximation on the large scales
of interest here. 

Second, we use an improved version of the hybrid simulation technique of Zahn et al. (2007), which
is essentially a Monte-Carlo implementation of the analytic model developed by Furlanetto et al.
(2004). This technique has the advantage of being extremely fast, while maintaining accuracy.
In comparison to the scheme described in Zahn et al. (2007), our present scheme is improved in
several ways. First, we use 2nd-order Lagrangian Perturbation Theory (2LPT) to generate realizations of the density field (Crocce et al. 2006) during reionization (as in Lidz et al. 2007b),
rather than generating Gaussian random fields. This allows us to incorporate quasi-linear effects.
Next, we use a scheme similar to that of Mesinger \& Furlanetto (2007a) to predict the
halo distribution from an initial, linear displacement field.

\subsection{Basic Simulated Signal}
\label{sec:sim_sig}

\begin{figure*}
\bc
\includegraphics[width=17.5cm]{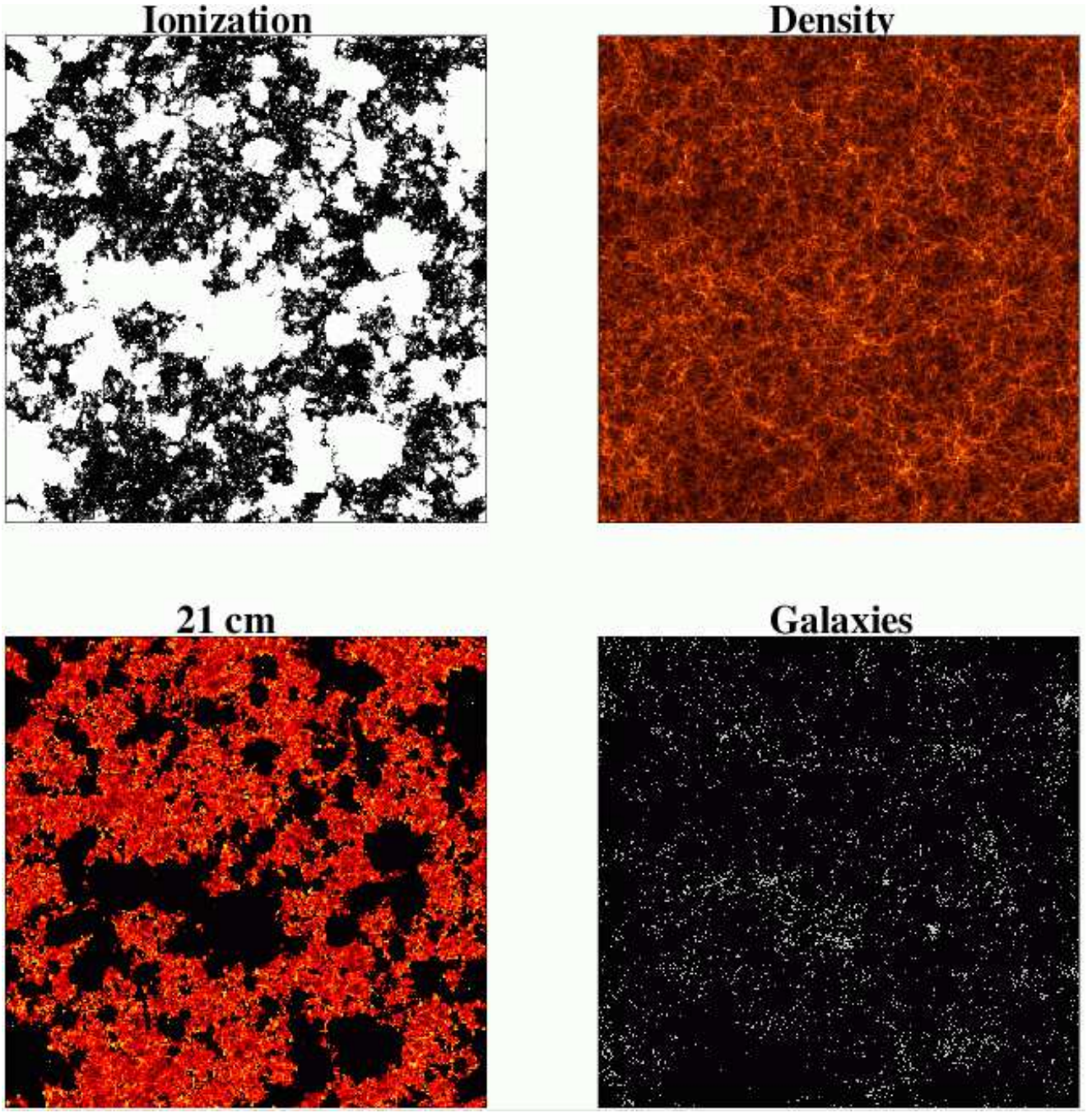}
\caption{Simulated maps of the density, halo, ionization, and 21 cm fields.
Each map is $130$ Mpc/$h$ on a side and is drawn from a simulation
snapshot at $z=7.32$ at which point $\avg{x_i} = 0.54$ in our model.
The density, ionization, and 21 cm maps are each $1$-cell thick ($0.25$
Mpc/$h$), while the halo field is from a $60$-cell ($15$ Mpc/$h$) wedge.
On large scales, the bright regions in the overdensity map tend to have
more halos, be ionized, and be dim in 21 cm. The correspondence between
the bright regions in the halo field, and the dim regions in the 21 cm
field, is the signal we characterize and quantify in this paper.
}
\label{fig:maps}
\ec
\end{figure*}

Let us now examine the main features of the simulated signal. To begin with, we consider the
McQuinn et al. (2007b) simulations, and focus on a model in which all
halos down to the atomic cooling mass contain sources with an ionizing 
luminosity proportional to host halo mass. Further, we assume that
all halos above $M_{\rm g, min} = 10^{10} M_\odot$ contain galaxies
detectable by our hypothetical survey.
In what follows, this prescription for the ionizing sources and the 
minimum detectable host halo mass constitutes our fiducial model. We denote
the minimum detectable host halo mass by $M_{\rm g, min}$, and the minimum host halo
mass for the ionizing sources by $M_{\rm x, min}$. 
Presently we consider
a simulation snapshot at $z = 7.32$, at which point the filling factor of 
ionized regions in our model is $\avg{x_i} = 0.54$.

It is illuminating to inspect the simulated fields visually before calculating
their detailed statistics. In Figure \ref{fig:maps} we show narrow 
slices through
our simulated density, halo, ionization, and 21 cm fields. Here one
can clearly see that the bright regions in the halo map correspond
to dim regions in the 21 cm map, while dim regions in the halo map
correspond to bright regions in the 21 cm map. This {\em anti-correlation}
is the signal we characterize and calculate in the present paper.
As one can see from the panels of Figure \ref{fig:maps}, the
anti-correlation arises because galaxies are more abundant in large scale 
overdense regions,
which hence ionize before typical regions. As a result, the overdense
regions contain less neutral hydrogen during reionization, and emit more
dimly in 21 cm than typical regions, while containing more galaxies
(see also Wyithe \& Loeb 2007).

\begin{figure}
\bc
\includegraphics[width=9.2cm]{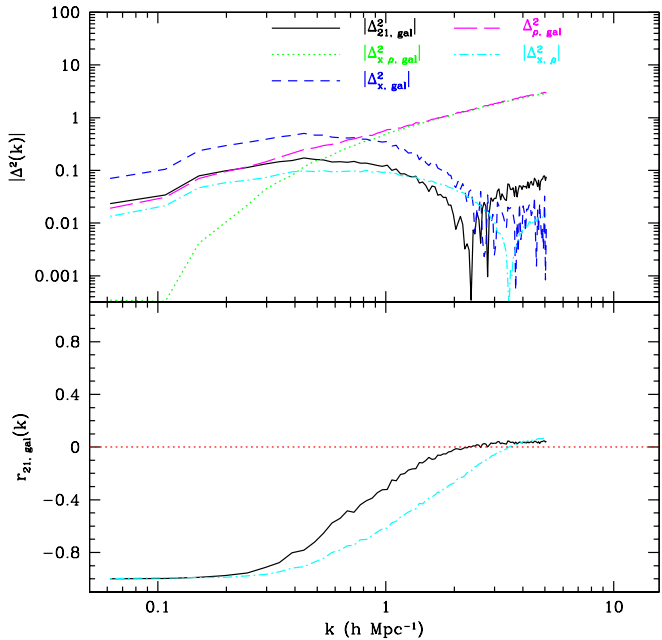}
\caption{The 21 cm-galaxy cross power spectrum and its constituent terms.
The signal is shown for our fiducial model at $\avg{x_i} = 0.54$.
{\em Top panel}: The absolute value of the 21 cm-galaxy cross power
spectrum (black solid line). The blue dashed line
shows the {\em x-gal} cross power spectrum, the
magenta long-dashed line shows the {\em $\rho-gal$} cross power spectrum,
and the green dotted line shows the {\em three-field} term. On small scales
the {\em three-field} and {\em $\rho-gal$} cross power spectra cancel each other
out rather closely. For contrast, we also show the cross power spectrum
between the neutral hydrogenic fraction and the density field (cyan dot-dashed
line).
{\em Bottom panel}: The cross correlation coefficient between the 21 cm
and galaxy fields as a function of wavenumber. The cyan dot-dashed line indicates the
cross-correlation coefficient between the neutral hydrogenic and density fields.
The red dotted line indicates
zero correlation coefficient. The sign of the signal in the top panel
can be inferred from the correlation coefficient shown here.
}
\label{fig:fid_decomp}
\ec
\end{figure}

In order to quantify these visual impressions, we calculate and
show the
21 cm-galaxy cross power spectrum in Figure \ref{fig:fid_decomp}. The
{\em top panel} shows the absolute value of the 21 cm-galaxy cross power spectrum, 
as well as the individual terms of Equation (\ref{eq:p21_gal_cross}). The {\em bottom
panel} shows the cross correlation coefficient between the two fields, 
$r(k) = P_{\rm 21, gal}(k)/\left[P_{\rm 21}(k) P_{\rm gal}(k)\right]^{1/2}$. In 
estimating the cross-correlation coefficient here and throughout this paper, we 
subtract shot-noise from the galaxy
power spectrum (before calculating $r(k)$) assuming that it is Poisson -- i.e., we assume 
$P_{\rm shot} = 1/n_{\rm gal}$, where
$n_{\rm gal}$ is the abundance of halos above $M_{\rm g, min}$. 

The figure reveals several interesting features of the signal. On large scales
the 21 cm field is anti-correlated with the galaxy field. As explained and
visualized in Figure \ref{fig:maps}, this occurs because galaxies form
first, and ionize their surroundings, in overdense regions.
On the other hand, on small scales the 21 cm and galaxy fields are 
roughly un-correlated. We can understand this by examining the small-scale behavior
of the constituent terms, as shown in the {\em top panel}. The cross power spectrum
between neutral hydrogen fraction and galactic density ($\Delta^2_{\rm x, gal}(k)$, the
{\em x-gal} term) turns over on small scales, as 
indicated by the blue-dashed line. This behavior is naturally similar to that of the
density-ionization cross power spectrum, which turns over on scales smaller than 
the size of the H II regions during 
reionization (Furlanetto et al. 2004, Zahn et al. 2007). The correlations die off on sub-bubble
scales because the 
entire interior of each H II region is highly ionized, irrespective of the interior 
density and galaxy fields.
For comparison, we additionally plot the cross 
power spectrum between neutral hydrogen fraction and matter density. This resembles
the cross power spectrum between neutral hydrogen fraction and galactic density, but
it turns over on slightly smaller scales.  As we explore further in \S \ref{sec:sources} and
\S \ref{sec:min_mass},
the turnover is on smaller scales owing to ionized bubbles around low mass halos, which 
host galaxies below the detection threshold of our hypothetical galaxy survey.

The cross power spectrum
between the density field and the galaxy field is shown by the long-dashed magenta
line. Note the very strong clustering of these rare galaxies: the cross power spectrum
has an amplitude of unity, $\Delta^2_{\rm \rho, gal}(k) \sim 1$, on a scale of
$k \sim 1.8 h$ Mpc$^{-1}$. On the same scale the amplitude of the matter power 
spectrum, $\Delta^2_{\rm \rho, \rho}(k)$, is a factor of $\gtrsim 7$ smaller. Hence
even though dark matter clustering is quasi-linear on relevant scales, the clustering
of detectable host-halos may be quite non-linear on the same scales. 

Finally, let us
examine the {\em three-field} term, $\Delta^2_{\rm x \rho, gal}(k)$. This term is negative
in our calculations and appears to closely cancel out the {\em $\rho-gal$} cross power 
spectrum on small scales. Owing to this cancellation, the shape of the 21 cm-galaxy
cross power spectrum closely mimics that of the {\em x-gal} cross
power spectrum. 
The 21 cm-galaxy cross correlation may then offer a relatively direct tracer of bubble
growth during reionization: it traces the {\em x-gal} term, which turns over
on scales smaller than that of the H II regions around the minimum mass detectable galaxies.
We examine this further in \S \ref{sec:evol_sig}, but we first pause to
consider the {\em three-field} term more closely.

In order to understand why the {\em three field} and {\em $\rho-gal$} terms 
cancel each other on small scales, it
is helpful to combine the two terms into a single one, and consider the two-point
correlation function rather than the power spectrum. Here we use similar reasoning
to that of Lidz et al. (2007a); see their \S 3.2.
The two terms are combined as: 
\beqa
\Delta^2_{\rm \rho, gal}(k) + \Delta^2_{\rm x \rho, gal}(k) \propto
\rm{F.T.}\left[\avg{x(1) \delta_\rho (1) n_g(2)}\right].
\label{eq:combine_cancel}
\eeqa
Here $1$ and $2$ indicate spatial positions $\x_1$ and $\x_2$, respectively,
while $\rm{F.T.}$ refers to a Fourier transform. This equation follows from
expanding $x(1)$ on the right-hand side of the equation as 
$x(1) = \avg{x} (1 + \delta_x(1))$, and using 
$\avg{\left[1+\delta_x(1)\right] \delta_\rho(1) n_g(2)} = 
\avg{\delta_\rho(1) n_g(2)} + \avg{\delta_x(1) \delta_\rho(1) n_g(2)} $.

We can write the above two-point function as: 
\beqa
\avg{x(1) \delta_\rho(1) n_g(2)} = \int dx(1) d\delta_\rho(1) dn_g(2) \times \nonumber \\
 x(1) \delta_\rho(1) n_g(2) P\left[x(1),\delta_\rho(1) | n_g(2)\right] 
P\left[n_g(2)\right].
\label{eq:twop}
\eeqa
Provided we consider separations much smaller than the size of the H II regions,
a pair of points $(1)$ and $(2)$ will mostly be either each within the same ionized
bubble, or both outside an ionized bubble. If each point is within a bubble then the
pixel at position $(1)$ is ionized, $x(1)=0$, and this gives no 
contribution to the two-point function of Equation (\ref{eq:twop}). On the other hand,
spatial points outside of bubbles do not contain detectable galaxies in this
model (although see \S \ref{sec:sources} for alternate cases), $n_g(2)=0$, and again
yield vanishing contributions to the two-point function of Equation (\ref{eq:twop}). 
The two-point function of Equation (\ref{eq:twop}) must hence vanish on small scales in
our fiducial scenario, which explains the cancellation between the {\em $\rho-gal$} and
{\em three-field} terms seen in Figure \ref{fig:fid_decomp}. The ionization field 
is a `mask' that surrounds {\em each}
galaxy in this model, eliminating the two-point function of Equation \ref{eq:twop} on
small scales. 

\subsection{Hybrid Simulations}
\label{sec:hybrid}

In addition to the full radiative transfer simulations of McQuinn et al. (2007a, 2007b),
we perform some calculations with the rapid hybrid scheme of
Zahn et al. (2007), Mesinger \& Furlanetto (2007a). We use hybrid simulations with 
two different boxsizes in this
work: one has $L_{\rm box} = 70$ Mpc/$h$, while the other
has $L_{\rm box} = 130$ Mpc/$h$. The density, halo, ionization, and 21 cm fields in each
simulation are tabulated on $512^3$ grid cells. In order to locate the halos using the
scheme of Mesinger \& Furlanetto (2007a), we
employ a grid of $1200^3$ cells in each simulation.  
The smaller box calculation has higher resolution -- resolving halos down to 
$M_{\rm min} \sim 10^8 M_\odot$ -- allowing us to accurately identify
halos with mass around the atomic cooling mass.  
The larger
box has coarser mass resolution, $M_{\rm min} \sim 10^9 M_\odot$, but 
better captures the small-$k$ 21 cm-galaxy cross spectrum. We refer the reader
to Zahn et al. (2007), Mesinger \& Furlanetto (2007a) for a detailed description and 
tests of the hybrid scheme. Here we briefly show that estimates of the 21 cm-galaxy
cross power spectrum from our hybrid simulations agree well with those from the full
radiative transfer simulations. 

\begin{figure}
\bc
\includegraphics[width=9.2cm]{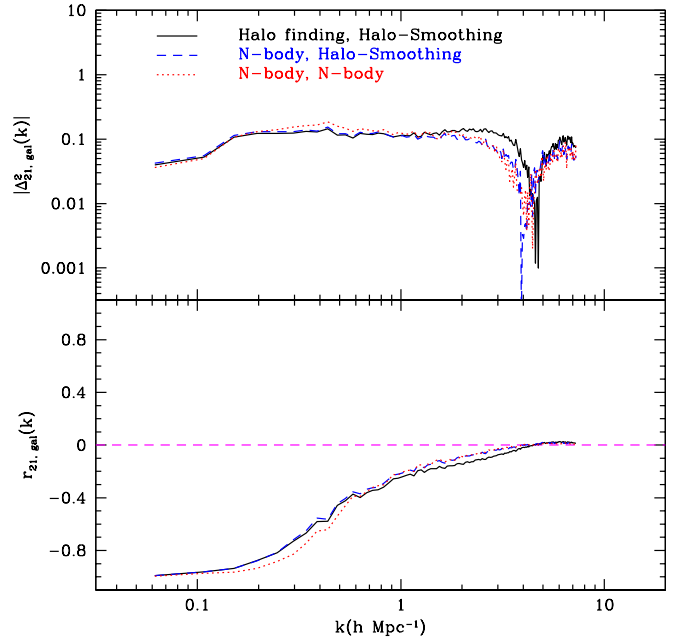}
\caption{Comparison between the hybrid and simulated 21 cm-galaxy cross 
spectrum. {\em Top panel}: The absolute value of the 21 cm-galaxy cross power spectrum.
{\em Bottom panel}: The cross-correlation coefficient between the two fields. In each panel,
the red dotted line shows the results from the full radiative transfer simulation. The black
solid line shows results from cross-correlating the hybrid 21 cm field with the hybrid halo
field. The blue dashed line shows the cross-correlation between the hybrid 21 cm field, and
the halo field from the N-body simulation. 
}
\label{fig:rco_21cm_gal_hf}
\ec
\end{figure}

In order to do this comparison, we use the initial conditions from the McQuinn et al. (2007b)
N-body simulation, and generate the halo field and ionization field using our hybrid scheme.
For the purposes of this comparison, in each of our hybrid and radiative transfer calculations,
we include ionizing sources only in halos that are
well resolved by the N-body simulation, with $M_{\rm x, min} \geq 8 \times 10^9 M_\odot$. That 
is, 
here we do not add low mass halos into the radiative transfer simulation with the appropriate 
statistical properties as in McQuinn et al. (2007a) and other sections of this paper. We limit 
our comparison to masses above $8 \times 10^9 M_\odot$
because these are the halos directly resolved in our N-body simulation, before small mass
halos are included statistically. 
We cross-correlate the resulting 21 cm field with all halos above our fiducial choice of 
$M_{\rm g, min} = 10^{10} M_\odot$.
The results of this comparison are shown in Figure \ref{fig:rco_21cm_gal_hf}, for outputs
with $\avg{x_i}$ just slightly below $\avg{x_i} = 0.5$. 

The agreement
between the hybrid and full radiative transfer calculations is quite good. In order to check how
much of the small difference between the two calculations comes from differences in the 21 cm
field, and how much from differences in the halo fields, we cross-correlate the hybrid
21 cm field with the simulated halo field (blue dashed lines). 
Differences in the simulated and hybrid halo fields seem to be important on small scales, while 
differences between the 21 cm fields in the two calculations lead to most of the difference 
on large scales.  
Regardless, the hybrid calculations agree well with the
full radiative transfer ones, and provide a useful means to estimate the 21 cm-galaxy cross
spectrum rapidly.

\section{Redshift Evolution of 21 cm-galaxy cross power spectrum}
\label{sec:evol_sig}

Now that we have introduced our simulation tools and 
understand the basic 21 cm-galaxy cross power spectrum signal,
let us examine its dependence on redshift and ionization fraction. 
How does the signal evolve as the filling factor of H II regions, and their characteristic
size, increase? To address this, we calculate the 21 cm-galaxy cross power spectrum 
from our radiative transfer simulations, considering a wide range of redshifts in 
order to span most of the reionization epoch. We start by adopting our fiducial
model with $M_{\rm g, min} = 10^{10} M_\odot$ at each redshift for simplicity.

\begin{figure}
\bc
\includegraphics[width=9.2cm]{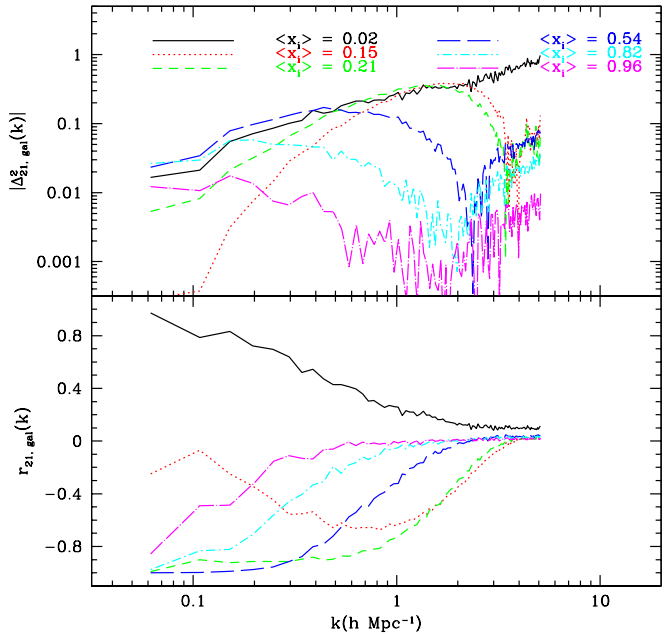}
\caption{Redshift evolution of the 21 cm-galaxy cross power spectrum
in our fiducial model. {\em Top panel}: The absolute value
of the 21 cm-galaxy cross power spectrum at different ionization
fractions/redshifts. The redshifts and ionization fractions shown 
are $(\avg{x_i}, z) = (0.02, 11.46); (0.15, 8.76);
(0.21, 8.34); (0.54, 7.32); (0.82, 6.90);$ and $(0.96, 6.77)$. 
{\em Bottom panel}: The cross-correlation coefficient between the 21 cm
and galaxy fields as a function of wavenumber.  
}
\label{fig:cross_v_z}
\ec
\end{figure}

The results of this calculation are shown in Figure \ref{fig:cross_v_z}.
At early times, near the very beginning of the reionization process
($\avg{x_i} = 0.02$), the galaxy and 21 cm fields are
{\em positively} correlated on large scales. At this stage, galaxies are
extremely rare objects and are only just starting to ionize their surroundings.
The galaxies turn on first in large scale overdense regions, which contain
more matter and initially more neutral hydrogen than large scale underdense
regions. These regions 
hence glow more brightly in 21 cm emission and lead to a positive 
21 cm-galaxy cross-correlation on large scales, as shown by the black solid
line in Figure \ref{fig:cross_v_z}.

The galaxies
quickly ionize their overdense surroundings which consequently dim in
21 cm emission. On the other hand, the large scale underdense regions are
still mostly free of galaxies and roughly maintain their initial 21 cm brightness 
temperature. This leads to a brief period where the 21 cm-galaxy cross
correlation has low amplitude on large scales, as large scale overdense
regions dim in 21 cm emission and roughly equilibrate in brightness temperature
with large scale underdense regions 
(Furlanetto et al. 2004, Wyithe \& Morales 2007; Lidz et al. 2007b 
discuss a similar low-amplitude
epoch for the 21 cm power spectrum).
In our fiducial model, this `equilibration phase' occurs
when $\avg{x_i} \sim 0.15$, as shown by the red dotted line in the figure. This equilibration
epoch is relatively brief; the two fields quickly become anti-correlated on large scales.

A caveat to this discussion is that our calculations assume that the spin temperature
of the 21 cm transition is globally much larger than the CMB temperature. This approximation
will be inaccurate early in the reionization process (Pritchard \& Furlanetto 2007, 
Pritchard \& Loeb 2008), and spin
temperature fluctuations may
complicate the cross-correlation signal close to the equilibration phase. 
This
is beyond the scope of the present paper, but may modify our results at very early times, perhaps
when $\avg{x_i} \lesssim 0.1$ (Pritchard \& Furlanetto 2007).

More robust, and detectable in the near future, are our results during the bulk of
the reionization process, at which point the 21 cm and galaxy fields are anti-correlated.
Once the anti-correlation is established, its scale dependence varies with redshift and
ionization fraction.
This behavior is shown in the green short-dashed, blue long-dashed, cyan dot-dashed and
magenta dash-dotted lines which span model ionization fractions of 
$\avg{x_i} = 0.21$ -- $0.96$. 
As discussed in \S \ref{sec:cross}, this anti-correlation reflects the fact that galaxies 
turn on first in
overdense environments and ionize their surroundings.
As the ionized fraction increases, and the H II regions grow, the cross-correlation turns over 
on progressively larger scales.
This illustrates that the 21 cm-galaxy cross power spectrum 
provides a relatively direct probe of bubble growth during reionization.
 
We pause here to mention one slight caveat regarding our modeling of the small scale 
cross spectrum. The galactic
sources will themselves contain neutral hydrogen, a feature which is not properly included in our
calculations. (This leads to a 21 cm signal {\em after} reionization; see Wyithe \& Loeb 2008, Chang
et al. 2008.)  This neglected contribution should cause the cross spectrum to 
become positive on small 
scales. Since the signal
from the diffuse IGM is much stronger on relevant scales than this galactic contribution (see
Lidz et al. 2007b, their \S 2.3, for an estimate), we do not expect this
to confuse the determination of the bubble-size induced turnover.  

As remarked previously, the precise turnover scale depends on the minimum host halo mass
of the galaxies observed by our hypothetical survey. Roughly speaking, the turnover
is set by the size of ionized regions around detectable galaxy hosts, and is insensitive
to the size of ionized regions around fainter galaxies (see \S \ref{sec:min_mass}). 
In practice, the minimum detectable host mass -- which impacts the turnover scale -- 
may vary with redshift in a complicated
way, depending on the flux-limit of the survey and the correlation between luminosity
and halo mass. This will make the evolution of the 21 cm-galaxy cross spectrum more
complicated than the illustrative results of Figure \ref{fig:cross_v_z}, which are at fixed
minimum host mass. In a flux-limited survey, the turnover scale will generally evolve
{\em less} strongly with time: in this case, one detects only massive galaxies at early
times, which tend to reside in larger bubbles than average.
In order to disentangle the impact of varying minimum host mass and that
of varying bubble size and ionization fraction, one could cross-correlate the 21 cm
signal with galaxies of varying luminosity and
use the galaxy auto spectrum and luminosity function to help understand
the correlation between galaxy luminosity and host halo mass. As we show 
in \S \ref{sec:min_mass}, 
measuring
the turnover scale as a function of galaxy luminosity, allows one to determine
the characteristic size of ionized bubbles as a function of luminosity.

The 21 cm auto spectrum itself evolves as the filling factor of H II regions 
increases. Lidz et al. (2007b) explored how one might use the redshift evolution of the auto
spectrum to constrain the evolution of the H II region filling factor. The redshift 
evolution of the cross spectrum, as considered here, would provide a complementary and 
essentially independent means for constraining H II region growth. Ultimately, combining
the two measurements should provide a cross check on each measurement 
and increase constraining power. More important, the cross spectrum provides a much
more direct indicator of characteristic bubble size than the auto spectrum. By measuring the 
cross spectrum
in different galaxy luminosity bins, one can additionally determine how the bubble size depends
on galaxy luminosity, information which is not obtainable from the 21 cm auto spectrum alone. 
Note that the ionized regions form under the collective influence of many individual galaxies,
but one still expects a statistical trend of bubble size with galaxy luminosity: more luminous
galaxies tend to live in more massive halos, which inhabit larger overdensities, and are typically
surrounded by larger ionized regions. Measuring the turnover in the cross spectrum for different
galaxy luminosity bins offers a unique means of quantifying this trend. 
We will discuss the statistical power of several future surveys to detect the
cross spectrum evolution in \S \ref{sec:detectability}.

\section{Dependence on Ionizing Source Properties}
\label{sec:sources}

Now that we understand the basic features of the 21 cm-galaxy cross spectrum, we consider
variations around our fiducial model parameters.
First, let us examine how the signal depends on the properties
of the ionizing sources. Precisely which sources of light produce most of the photons
that reionize the IGM is highly uncertain. This depends on many poorly constrained quantities
such as the efficiency of star formation as a function of galaxy mass, the high redshift
stellar initial mass function (IMF), 
the fraction of ionizing
photons that escape host galaxies to ionize the IGM and its dependence on host mass, the
degree to which photoionization and supernova feedback suppress star formation in low mass halos,
and other factors (e.g. Furlanetto et al. 2006a). A promising route to constrain some of these 
uncertain parameters is to study
the differing impact these sources have on the surrounding IGM. Put simply, the IGM may provide
a valuable laboratory for studying the first luminous sources.
In this section we show that measurements of the 21 cm-galaxy cross spectrum may help constrain
ionizing source properties.

To explore this, let us start with a simple model and vary two of our model parameters. 
First, we vary $M_{\rm x, min}$, the minimum mass of halos that host sources contributing to
reionization. Next, we vary $M_{\rm g, min}$, the
minimum host mass of galaxies detectable by our hypothetical galaxy survey. We explore the
impact of varying these parameters using $70$ Mpc/$h$ hybrid simulations, each normalized to
$\avg{x_i} = 0.5$ at $z = 6.9$. 
To begin with, we fix the parameter $M_{\rm g, min}$ at $10^{10} M_\odot/h$ and
consider $M_{\rm x, min} = 10^8 M_\odot/h$, 
$10^9 M_\odot/h$,
and $10^{10} M_\odot/h$ respectively.\footnote{Note that we generally quote masses in units
of $M_\odot$, but here and in \S \ref{sec:min_mass} (owing to imperfect planning) we use $M_\odot/h$ 
units, and so the choice of 
$M_{\rm g, min} = 10^{10} M_\odot/h$ is slightly different than our fiducial choice of
$M_{\rm g, min} = 10^{10} M_\odot$.}
These are clearly simplified models, but they
suffice to illustrate the basic sensitivity of the signal to ionizing source properties. 
These models should approximate
scenarios in which photo-heating (Thoul \& Weinberg 1996, Navarro \& Steinmetz 1997, 
Dijkstra et al. 2004) or supernova feedback (e.g. Springel \& Hernquist 2003) limit the
efficiency of star-formation in small mass halos and diminish the contribution of 
these halos to reionization.

\begin{figure}
\bc
\includegraphics[width=9.2cm]{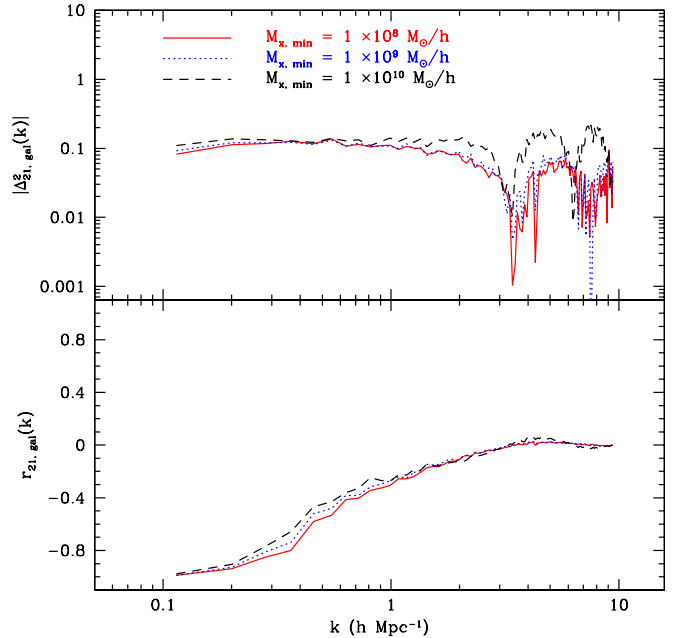}
\caption{The 21 cm-galaxy cross power spectrum for models of varying $M_{\rm x, min}$.
In each model curve the efficiency of the ionizing sources is adjusted to yield
$\avg{x_i} \sim 0.5$ at $z=6.9$, and the minimum detectable galaxy host has a mass
of $M_{\rm g, min} = 10^{10} M_\odot/h$. {\em Top panel}: The absolute value of the
21 cm-galaxy cross power spectrum. {\em Bottom panel}: The correlation coefficient
between the galaxy and 21 cm fields in each case. The dependence of the cross power
spectrum on the host halo mass of the ionizing sources is rather mild. 
}
\label{fig:rco_xmin}
\ec
\end{figure}

Varying $M_{\rm x, min}$ across the range shown in Figure \ref{fig:rco_xmin} 
only weakly  
influences the 21 cm-galaxy cross
spectrum. On large scales, the amplitude increases slightly with 
increasing $M_{\rm x, min}$,
since the bias of the ionized regions is larger for larger values of
$M_{\rm x, min}$. 
Note, however, that the small-scale turnover occurs at very similar scales
for each $M_{\rm x, min}$. This occurs because, in each case shown here, the minimum
detectable galaxy mass is larger than (or equal to) $M_{\rm x, min}$.
The cross-correlation is mostly
insensitive to the bubble sizes around smaller mass, undetectable sources. As alluded
to earlier, the {\em turnover in the cross spectrum depends on the bubble sizes around
galaxies above the minimum mass detectable by our hypothetical galaxy survey}, and is
mostly insensitive to the bubble sizes around lower mass hosts. Note that the auto spectra
of the ionization and 21 cm fields {\em do} depend on $M_{\rm x, min}$ (Furlanetto et al. 2006b,
McQuinn et al. 2007a, Lidz et al. 2007b) -- models with larger $M_{\rm x, min}$ have larger
bubbles (on average) at a given $\avg{x_i}$. However, it appears that the bubble sizes around high mass
galaxies (with $M \gtrsim M_{\rm g, min}$) change only slightly with increasing $M_{\rm x, min}$, 
and hence the turnover scale in the cross spectrum is insensitive to $M_{\rm x, min}$. We have
verified this explicitly by calculating the average ionization as a function of distance
around halos with $M=M_{\rm g, min}$, for each of $M_{\rm x, min} = 10^8 M_\odot/h$ and
$M_{\rm x, min} = 10^{10} M_\odot/h$. The ionization profiles around the massive halos are very
similar in these models, supporting our interpretation. 

Another possibility is that ionizing photons escape efficiently only from high mass 
galaxies,
and that low mass sources do not contribute to reionizing the IGM. This possibility 
is, in fact, suggested by the recent 
escape fraction simulations of Gnedin et al. (2008).
Even if low mass galaxies have a 
negligible
escape fraction, they may still form stars efficiently and 
be detectable at wavelengths longward of the hydrogen ionization edge. 
This scenario
produces an interesting signature in the 21 cm-galaxy cross power spectrum, provided 
one has
a galaxy survey capable of detecting the, presumably faint, sources in these low mass 
halos.
In order to explore this, we fix the minimum host halo mass of sources contributing to
the reionization of the IGM at $M_{\rm x, min} = 10^{10} M_\odot/h$ and calculate the
21 cm-galaxy cross spectrum with a galaxy survey probing sources in host halo masses
larger than each of $M_{\rm g, min} = 10^8, 10^9,$ and $10^{10} M_\odot/h$. 

\begin{figure}
\bc
\includegraphics[width=9.2cm]{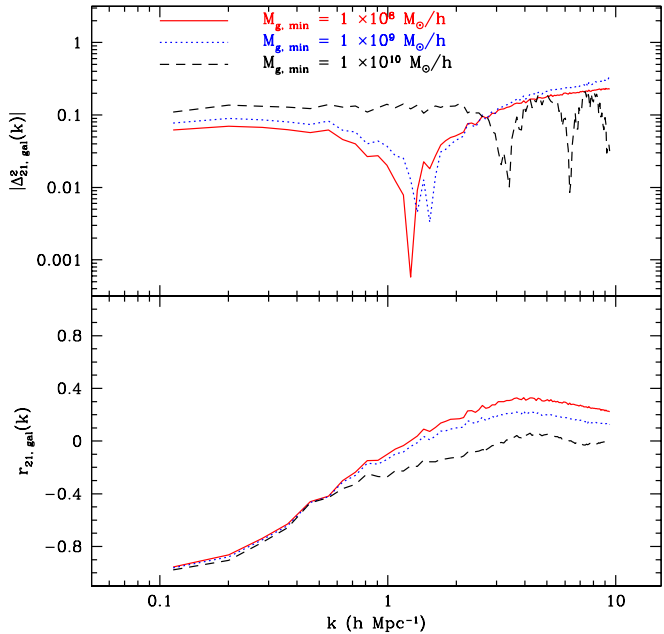}
\caption{21 cm-galaxy cross power spectra for models where sources in high
mass halos produce most of the ionizing photons. In each model, the minimum
host halo mass of sources that allow ionizing photons to escape into the IGM
is $M_{\rm x, min} = 10^{10} M_\odot/h$. {\em Top panel}: The absolute value
of the 21 cm-galaxy cross spectrum for galaxy surveys with minimum detectable 
host halo masses of $M_{\rm g, min} = 10^8, 10^9$ and $10^{10} M_\odot/h$. {\em Bottom panel}: The
correlation coefficient between the galaxy and 21 cm fields in each case. The cross spectrum
and correlation coefficient turn positive on small scales for cases in which the galaxy
survey detects sources with mass below the minimum mass ionizing source. 
}
\label{fig:rco_gmin}
\ec
\end{figure}

The results of this calculation are shown in Figure \ref{fig:rco_gmin}. On large scales,
the amplitude of the cross spectrum increases as one raises the minimum detectable host
halo mass. This increase simply owes to the usual increase in galaxy bias with increasing 
minimum host halo mass.
Perhaps more interesting, however, are 
the results on small scales when the minimum detectable
host halo mass is {\em lower} than the minimum ionizing source mass. In this case (see
the model curves with $M_{\rm g, min} = 10^8$ and $10^9 M_\odot/h$), the
cross spectrum turns over on larger scales than in the model in which 
$M_{\rm g, min} = M_{\rm x, min}$, it then reverses sign, and goes positive on small 
scales.
Detecting this behavior would indicate that ionizing photons escape only from massive 
host halos,
and not from lower mass hosts. An ambitious survey is needed to detect the faint 
sources in
low mass halos, and to detect the 21 cm-galaxy cross spectrum on small scales (see
\S \ref{sec:detectability}).
Nonetheless,
the proposed signature would provide an interesting indication of a 
small escape fraction from low mass galaxies. Moreover, this signature is relatively
direct -- any indication of a small escape fraction from low mass galaxies in the 21 cm auto
spectrum will be more subtle, and likely degenerate with other effects. Note, however, that
neutral hydrogen in the galaxies themselves may also result in a positive small-scale cross
spectrum (\S \ref{sec:evol_sig}), and it might be tricky to distinguish this from our
escape fraction scenario.

\begin{figure}
\bc
\includegraphics[width=9.2cm]{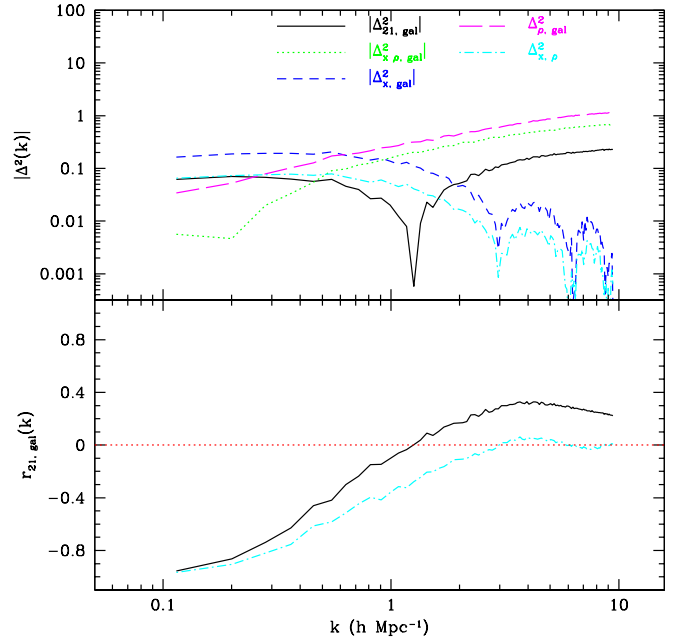}
\caption{Decomposition of the 21 cm-galaxy cross spectrum for a model in which the minimum
detectable galaxy host mass is below the minimum host mass for ionizing sources. Here
we show the decomposition for a 
model with $M_{\rm x, min} = 10^{10} M_\odot/h$ and $M_{\rm g, min} = 10^8 M_\odot/h$. 
{\em Top panel}: As in Figure \ref{fig:fid_decomp}, the absolute value of the 
21 cm-galaxy cross spectrum, as well as the {\em x-gal}, {\em $\rho-gal$} and
{\em three-field} terms. Additionally, we show the cross power spectrum between the
neutral hydrogen field and the density field itself. Unlike in our previous models, the
{\em three-field} and {\em $\rho-gal$} terms do not perfectly cancel each other on small
scales. This results because there are 
detectable galaxies outside
of ionized regions in this model. Consequently, the 21 cm-galaxy cross spectrum changes
sign around $k \sim 1 h$ Mpc$^{-1}$ and goes positive on small scales. 
{\em Bottom panel}: The cross-correlation coefficient between the 21 cm and galaxy fields,
as well as the cross-correlation between the neutral hydrogen and density fields.   
}
\label{fig:decomp_gmin}
\ec
\end{figure}

In order to understand this effect better, it is useful to calculate each of the
terms in Equation (\ref{eq:p21_gal_cross}) separately. In Figure \ref{fig:decomp_gmin} 
we examine each of these
terms for our model with $M_{\rm x, min} = 10^{10} M_\odot/h$ and 
$M_{\rm g, min} = 10^8 M_\odot/h$. In this case, the {\em three-field} and {\em $\rho-gal$}
terms do not cancel each other, unlike in our fiducial case (Figure \ref{fig:fid_decomp}).
This occurs because in this model low mass galaxies do not leak ionizing photons into
the IGM, and can hence reside outside of the ionized regions which -- in this model -- are formed 
only by sources residing
in higher mass halos. In this way, some low mass halos escape the `masking' effect of the
ionized regions (see \S \ref{sec:sim_sig} and Equation \ref{eq:twop}), and -- since 
these low mass galaxies are correlated with the underlying
density field -- produce a {\em positive} small scale 21 cm-galaxy cross power spectrum.

\section{The Impact of Lyman-limit Systems}
\label{sec:recombs}

Next we consider the impact of Lyman-limit systems on our 21 cm-galaxy cross spectrum 
calculations.
Once reionization is complete, most ionizing photons are absorbed in dense blobs of 
neutral gas 
known as Lyman-limit systems. Lyman-limit systems can also limit
the mean free path of ionizing photons and halt the growth of 
H II regions (Furlanetto \& Oh 2005, McQuinn et al. 2007a) during reionization itself, 
particularly 
towards the end of reionization.
The precise physical nature and abundance of these systems at high redshift is 
highly uncertain,
as is their role as photon sinks during reionization.
Lyman-limit systems may be especially numerous and have a strong effect if `mini-halos' --
halos with mass less than the atomic cooling mass --
manage to survive pre-heating prior to reionization (Oh \& Haiman 2003) and are abundant 
during
reionization (Haiman et al. 2001, Barkana \& Loeb 2002, Shapiro et al. 2004).

In order to quantify the impact of Lyman-limit systems on the 21 cm-galaxy cross 
spectrum, we use the hybrid simulation scheme of 
Zahn et al. (2007), generalized to include the recombination excursion-set 
barrier of Furlanetto \& Oh (2005). In order to capture the small-$k$ power spectrum --
where we expect the Lyman-limit systems to have the most impact -- we use the 
$L_{\rm box} = 130$ Mpc/$h$ hybrid simulation. In this section, for simplicity, our hybrid
simulation adopts the pure Press-Schechter ionization barrier
of Furlanetto et al. (2004), rather than the halo-smoothing algorithm (see Zahn et al.
(2007) for comparisons). Since the present work is the first to incorporate the recombination 
barrier into a hybrid simulation scheme, we briefly review this model here, but refer 
the reader to Furlanetto \& Oh (2005) for more details.  
The recombination barrier reflects the requirement that for an H II region to grow, the instantaneous rate of
photon production from the sources within the H II region must at least match the recombination rate 
of the ionized material inside the H II region.

In Furlanetto \& Oh (2005), the recombination rate is calculated using the model of
Miralda-Escud\'e et al. (2000). In this model at any given time, the interior of an H II region 
is ionized up to islands of small-scale
overdensity $\Delta_i$, above which the gas is neutral. The mean free path to ionizing photons is 
then
determined by the volume-filling factor of these overdense islands. In 
particular,
the {\em proper} mean free path to ionizing photons is given by:
\beqa
\lambda(z) = \lambda_0(z) \left[1 - F_v(\Delta_i)\right]^{-2/3},
\label{eq:lambda}
\eeqa
where $F_v(\Delta_i)$ denotes the volume-filling factor of regions with $\Delta < \Delta_i$, and
$\lambda_0(z)$ is a normalization factor, which is given by $\lambda_0(z) H(z) = 60$ km/s
in Miralda-Escud\'e et al (2000).
Here we leave $\lambda_0(z)$ as a free parameter to gauge the dependence of our results on the 
observationally and theoretically uncertain mean 
free path. The filling factor $F_v(\Delta_i)$ is computed using the gas density pdf in Miralda-Escud\'e et al. (2000). Similarly, the recombination rate for the ionized gas, in a region of large scale over-density $\delta$
ionized up to an overdensity $\Delta_i$,
is given by:
\beqa
A(\delta, \Delta_i) = \alpha_A n_e (1 + \delta) \int_0^{\Delta_i} d\Delta \Delta^2 P(\Delta).
\label{eq:cfac}
\eeqa
Here $\alpha_A$ denotes the case-A recombination coefficient for the ionized gas, which we assume to be
at $10^4$ K, $n_e$ denotes the mean electron density in the IGM, and $P(\Delta)$ is the gas density
pdf from Miralda-Escud\'e et al. (2000). We assume helium is mostly singly-ionized,
but not doubly-ionized, within the bubble interiors. The recombination rate 
formula assumes that the density pdf, $P(\Delta)$,
is independent of large-scale overdensity, $\delta$, which should be a good approximation for the large
scales relevant here (Furlanetto \& Oh 2005). 

With this formula for the recombination rate in hand, Furlanetto \& Oh (2005) write down an excursion set
barrier for a region of size $R$ and overdensity $\delta$ to overcome recombinations and be ionized by
interior sources. In our notation, this formula is:
\beqa
\zeta \frac{df_{\rm coll}(\delta, R)}{dt} > A(\delta, R),
\label{eq:rec_barr}
\eeqa
where $\zeta$ denotes the ionizing efficiency of the sources, $f_{\rm coll}$ denotes the collapse fraction
in halos above the minimum host halo mass, and $R$ is equated with the mean free path, $\lambda$,
which sets $\Delta_i$ through Equation (\ref{eq:lambda}), and $A(\delta, R)$ through Equation (\ref{eq:cfac}).
We implement this barrier, and apply it in a Monte Carlo fashion (Zahn et al. 2007, Mesinger \& Furlanetto 2007a), in conjunction with the normal
Furlanetto et al. (2004) barrier. 
This barrier effectively prohibits ionizing photons from propagating long distances, as regulated by the parameter $\lambda_0(z)$, and decreases the level of ionization fluctuations on large scales.

\begin{figure}
\bc
\includegraphics[width=9.2cm]{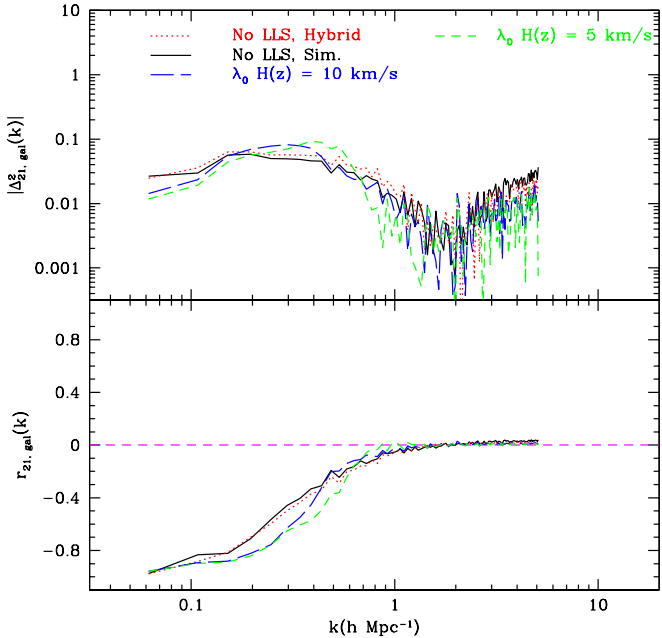}
\caption{Dependence of the 21 cm-galaxy cross power spectrum on the 
abundance of Lyman-limit systems. {\em Top panel}: The absolute value of the
cross spectrum. {\em Bottom panel}: The cross-correlation coefficient.
The black solid line shows the cross spectrum in our
fiducial model at $\avg{x_i} = 0.8$ ($z=6.90$). The red dotted line shows an equivalent
model from our hybrid simulation scheme. The green short-dashed and blue long-dashed lines
show cross spectrum calculations for models with abundant Lyman-limit systems (see text).
The recombination barrier scales (see text) for the models with 
$\lambda_0(z) H(z) = 5$ km/s, and $\lambda_0(z) H(z) = 10$ km/s are 
$R_{\rm rec, barr} = 4, 7$ Mpc/$h$ respectively.
The Lyman-limit systems force the cross spectrum to turn-over towards large scales, but the
effect is relatively mild. 
}
\label{fig:rco_recs}
\ec
\end{figure}

In order to see how this can impact the 21 cm-galaxy cross power spectrum, we calculate 
the signal
with hybrid simulations of varying $\lambda_0(z)$. In particular, we 
consider $\avg{x_i} = 0.8$ and
vary $\lambda_0(z)$ over the range $\lambda_0(z) H(z) = 5-60$ km/s. We span here a 
rather broad range of models, which is appropriate given our limited observational
constraints on the mean free path to ionizing photons at high redshift. The results of 
calculations with no Lyman-limit systems, and each of $\lambda_0(z) H(z) = 5$ and
$10$ km/s,
are shown in Figure \ref{fig:rco_recs}. Since $\lambda_0(z) H(z) = 5$ km/s is $1/12$th of
the fiducial Miralda-Escud\'e et al. (2000) value, our values represent rather extreme
choices for the mean free path.
A more meaningful characterization than
$\lambda_0(z)$, is the characteristic scale where the recombination barrier
(Equation \ref{eq:rec_barr}) crosses the usual Furlanetto et al. (2004) barrier 
(see Furlanetto \& Oh 2005). We 
denote the scale where the two barriers cross by $R_{\rm rec, barr}$.
For models with $\lambda_0(z) H(z) = 5, 10$ km/s (and $\avg{x_i} = 0.8$, $z=6.90$),
the barriers cross at respective radii of $R_{\rm rec, barr} = 4$ and $7$ Mpc/$h$.
Note that, for our choice of model parameters, the recombination barriers are 
not so steep, and so some fraction of 
points do manage to cross the barriers on smoothing scales roughly twice as large as 
$R_{\rm rec, barr}$.   
Comparing the red and black model
curves, we see that the hybrid scheme accurately captures the 21 cm-galaxy cross power
spectrum signal from our full radiative transfer simulations in the no Lyman-limit system
case as in Figure \ref{fig:rco_21cm_gal_hf}. 

The model curves with $\lambda_0(z) H(z) = 10$ km/s 
($R_{\rm rec, barr} = 7$ Mpc/$h$) and $\lambda_0(z) H(z) = 5$ km/s 
($R_{\rm rec, barr} = 4$ Mpc/$h$)
illustrate that decreasing the mean free path forces the 21 cm-galaxy cross power
spectrum to turn over towards large scales, rather than simply flattening out on large
scales as in our fiducial model.\footnote{In our fiducial model, the 21 cm-galaxy cross power
spectrum should turn over on some scale larger than that of our simulation box. Note however
that foreground contamination may make such scales inaccessible to future 21 cm observations
(McQuinn et al. 2006).} The cross spectrum
develops a more well defined characteristic scale as the mean free path decreases. 
This trend results because decreasing the mean free path limits the formation
of very large H II regions which in turn reduces the amount of large scale cross power.
Although the figure illustrates a clear trend of decreasing large scale power with
decreasing mean free path, we caution that there is no simple one-to-one correspondence
between mean free path and H II region size. As a specific illustration of the
distinction between the mean free path and H II region size, consider the 
post-reionization IGM. In the post reionization IGM, essentially the entire volume of 
the IGM is ionized, and the bubble size hence
infinite, while the mean free path is still finite. 

Note that although the figure shows a clear trend of decreasing large scale power with
decreasing mean free path, the dependence on the abundance of Lyman-limit systems 
is rather weak. This may
preclude strong constraints on the abundance of Lyman-limit systems from future measurements
of the 21 cm-galaxy cross power spectrum. On the other hand, it implies that the 
mean free path to ionizing photons is not a very important factor in our modeling
of the cross spectrum. This should allow us to robustly constrain other parameters
from future measurements, in spite of our ignorance of the high redshift mean free path.

\section{Dependence on Galaxy-Selection Technique} 
\label{sec:selection}

In this section, we consider the dependence of the 
21 cm-galaxy cross power spectrum on the manner in which the galaxies are selected.
Thus far we have calculated the 21 cm-galaxy cross power spectrum by cross-correlating our
21 cm field with all simulated halos above some minimum detectable halo mass cut. In 
other
words, we assume that each simulated dark matter halo contains one luminous galaxy, and that the
flux limit of our hypothetical galaxy survey corresponds precisely to a minimum host halo mass.
This is clearly a vast simplification, and so it is important to explore the signal's sensitivity to
the minimum mass cut. Note, however, that since we consider only scales much larger than
the halo virial radius, we are {\em not} sensitive to the distribution of galaxies within
each host halo (e.g. Scoccimarro et al. 2001).
Another important effect is that galaxies selected on the basis of Ly-$\alpha$
emission will have a different 21 cm-galaxy cross-correlation than galaxies selected by, for 
example, 
the Lyman-break technique. We presently explore the sensitivity of our results to the {\em type} of
galaxy selected by our hypothetical survey.

\subsection{Minimum Detectable Mass}
\label{sec:min_mass}

Let us first fix the population of galaxies responsible for reionizing
the IGM, and the resulting ionization field, while varying the minimum host halo mass containing
galaxies detectable by our hypothetical survey.
We explored this issue somewhat already in \S \ref{sec:sources}, but there we focused
on scenarios in which ionizing photons do not escape from low mass galaxies -- i.e., 
cases where $M_{\rm g, min} < M_{\rm x, min}$. Here we focus on models in which
ionizing photons manage to escape from low mass halos, yet such sources are too faint
to be detectable by our hypothetical galaxy survey. In other words, we consider cases
where $M_{\rm g, min} > M_{\rm x, min}$. This is likely the more relevant case
for first generation surveys where it will be difficult to detect the presumably
faint galaxies that reside in low mass halos.

Another point is that it is unlikely that {\em all} halos above some $M_{\rm g, min}$
host galaxies that actively produce detectable photons at any given instant of time. 
In other words, the `duty cycle' -- the fraction of halos above a given mass which
contain galaxies actively radiating at a particular time -- is likely less
than unity. 
As quantified below, varying the minimum detectable host mass impacts the mean 
21 cm-galaxy cross power 
spectrum, but
reducing the duty cycle of detectable galaxies does not by itself change the average cross power
spectrum signal. This is because galaxy bias is independent of duty cycle, provided that 
the duty
cycle is itself independent of mass for halos above the minimum detectable host mass. 
Decreasing the
duty cycle instead increases the level of Poisson fluctuations in the galaxy abundance, 
which
increases the cross spectrum {\em variance} -- and makes the cross spectrum more difficult to
detect (Furlanetto \& Lidz 2007, \S \ref{sec:detectability}) -- while preserving 
the {\em average} cross power spectrum. 
Presently, we focus on how the minimum host halo mass impacts the mean signal, and defer
a discussion of the signal variance to \S \ref{sec:detectability}.
Of course our assumption that the duty cycle is independent of host mass may be too simplistic and
modifying this assumption may impact our results in detail. Our simple model should, however, suffice 
to 
illustrate the basic sensitivity to host halo mass. Moreover, in practice one can constrain
the run of duty cycle with halo mass from the observed galaxy luminosity function and
galaxy-galaxy auto power spectrum, which can then inform models for the 21 cm-galaxy
cross spectrum.

\begin{figure}
\bc
\includegraphics[width=9.2cm]{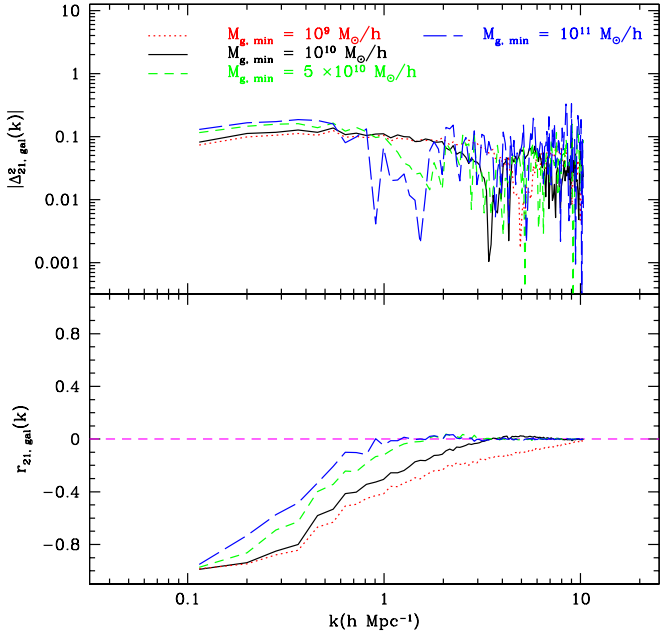}
\caption{Dependence of 21 cm-galaxy cross power spectrum on the minimum
detectable galaxy mass. {\em Top panel}: The cross power spectrum
for a survey that detects all galaxies in halos of mass larger
than $M_{\rm g, min} = 10^9 M_\odot/h$,
$M_{\rm g, min} = 10^{10} M_\odot/h, 5 \times 10^{10} M_\odot/h$ and 
$10^{11} M_\odot/h$ respectively. For each model curve, we fix the minimum
host mass of the ionizing sources at $M_{\rm x, min} = 10^8 M_\odot/h$. 
{\em Bottom panel}: The correlation coefficient between the
galaxy and 21 cm fields in each case.
}
\label{fig:rco_galmass}
\ec
\end{figure}

The results of varying the minimum detectable host halo mass are shown in Figure
\ref{fig:rco_galmass}. Here we use the $70$ Mpc/$h$ hybrid simulation, fix
$M_{\rm x, min} = 10^8 M_\odot/h$ (just a little above the atomic cooling mass), and vary
$M_{\rm g, min}$ from $10^9 M_\odot/h$ to $10^{10} M_\odot/h, 5 \times 10^{10} M_\odot/h$,
and $10^{11} M_\odot/h$. On large scales one sees the usual increase in the 
amplitude of the cross spectrum as $M_{\rm g, min}$ increases. On small scales, the cross spectrum
turns over on progressively smaller scales as $M_{\rm g, min}$ decreases, and the 
cross-correlation starts
to sample the small bubbles around the lower mass halos. This is a continuation of
the behavior seen in Figure \ref{fig:rco_xmin}. It illustrates that the turnover scale needs
to be interpreted with caution, since it is sensitive to $M_{\rm g, min}$. The dependence of
turnover scale on luminosity is very interesting, however; examining it amounts to a measurement of
the characteristic bubble size around galaxies of varying luminosity. In order to best constrain
this dependence one needs a galaxy survey with a sufficiently large dynamic range in 
luminosity, and 
one needs
to examine the luminosity dependence of the galaxy luminosity function, auto spectrum and 
cross spectrum. This also highlights the scientific benefit of measuring the 21 cm-galaxy cross
spectrum, as it is impossible to determine the luminosity dependence of the bubble 
size distribution
from the 21 cm auto spectrum alone.

\subsection{Lyman-alpha Selected Galaxies}
\label{sec:laes}

A successful approach for finding high redshift galaxies is to search for Ly-$\alpha$ emission,
which is frequently strong in young galaxies (Partridge \& Peebles 1967). There are numerous
existing and planned Ly-$\alpha$ emitter (LAE) surveys (e.g. Rhoads et al. 2004, Kashikawa et al. 2006, Stark et al. 2007), with the Subaru telescope currently providing the largest high redshift
sample, consisting of $\sim 58$ photometric LAEs at $z = 6.5$ discovered 
in a $\sim 30^{'} \times 30^{'}$
field (Kashikawa et al. 2006).  LAE surveys have an advantage over high redshift Lyman break
surveys in that they target narrow wavelength intervals, in between strong night sky background lines, in
search of strong emission lines. This allows one to detect galaxies that are unobservable by Lyman
break selection owing to the strong night sky background at the relevant wavelengths; sizable Lyman-break
galaxy catalogues likely await a widefield, near-infrared instrument in space or 30-meter class telescopes
on the ground.
Existing LAE surveys and their extensions 
hence likely provide the first opportunity to detect the 21 cm-galaxy
cross power spectrum, particularly if the IGM is partly neutral at $z \sim 6.6$ (Wyithe \& Loeb 2007,
Furlanetto \& Lidz 2007, \S \ref{sec:detectability}).

To this end, we would like to consider the cross correlation between 21 cm and {\em Ly-$\alpha$ selected
galaxies}. In contrast to galaxies selected via, for example the Lyman break or H-$\alpha$, the abundance 
of observable Ly-$\alpha$ selected galaxies will be modulated by the presence of neutral hydrogen, impacting
their clustering (Furlanetto et al. 2006c, McQuinn et al. 2007a, 2007b, Mesinger \& Furlanetto 2007b, 2007c) 
and the 21 cm-galaxy cross power
spectrum. This modulation occurs because damping wing absorption extinguishes the Ly-$\alpha$ line
for sources sufficiently close to the edge of an H II region (Miralda-Escud\'e 1998), where there
is an adjacent column of neutral hydrogen. This means Ly-$\alpha$ selected galaxies will lie towards
the center of large-ish, $R \gtrsim 1$ proper Mpc, H II regions 
(Furlanetto et al. 2006c, McQuinn et al. 2007a, 2007b, Mesinger \& Furlanetto 2007b, 2007c). Owing to
this, and because observable galaxies will have larger masses after Ly-$\alpha$ selection, the
clustering of Ly-$\alpha$ selected galaxies should increase as such galaxies are detected at earlier
and earlier stages of reionization.
Thus far our mock galaxies have been {\em uniformly selected} -- i.e., not modulated by the presence
of neutral hydrogen. While this is appropriate for Lyman break selected galaxies, it is incorrect for
Ly-$\alpha$ selected galaxies before reionization completes.

In order to examine the impact of Ly-$\alpha$ selection on the 21 cm-galaxy cross spectrum, we compute
the damping wing optical depth, $\tau_D$, towards each of our target halos. For simplicity, we calculate only
the damping wing optical depth at line 
center (see e.g. Equations (1) and (2) of Mesinger \& Furlanetto 2007c) and do not model resonant 
absorption (see McQuinn et al. 2007b, Mesinger \& Furlanetto 2007b, Dijkstra et al. 2007 for discussion).  Assuming that
each source's luminosity is proportional to its host halo mass, and adopting our fiducial choice
of $M_{\rm g, min} = 10^{10} M_\odot$, our Ly-$\alpha$ survey detects sources with $M \rm{exp}[-\tau_D] \geq
M_{\rm g, min}$.

\begin{figure}
\bc
\includegraphics[width=9.2cm]{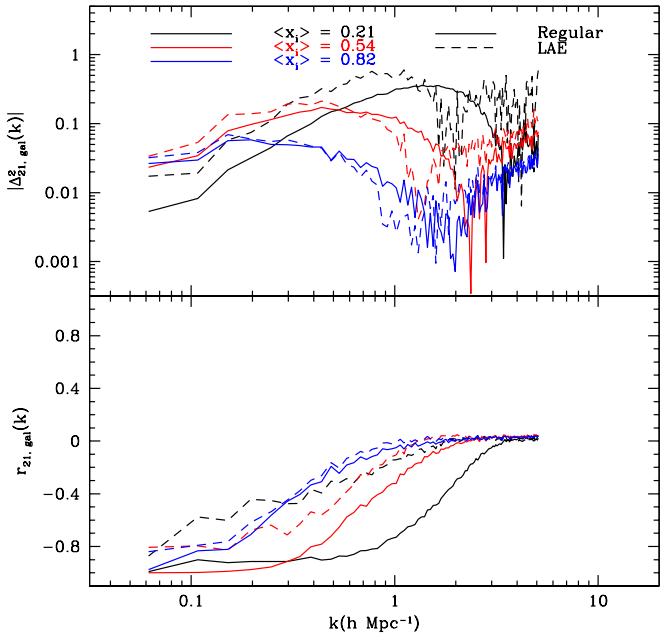}
\caption{The 21 cm-galaxy cross power spectrum for Ly-$\alpha$ selected
galaxies. {\em Top panel}: The 21 cm-galaxy cross power spectrum for each
of Ly-$\alpha$ selected galaxies (dashed lines) and `all galaxies' (solid lines). 
We show results at $(\avg{x_i}, z) = (0.21, 8.34), (0.54, 7.32), (0.82, 6.90)$, and
$M_{\rm g, min} = 10^{10} M_\odot$.
{\em Bottom panel}: The cross-correlation
coefficient between the 21 cm and galaxy fields for the model curves in the top panel. 
}
\label{fig:rco_lae}
\ec
\end{figure}

With mock Ly-$\alpha$ selected galaxy catalogues in hand, we calculate the cross 
power spectrum
of these galaxies with the 21 cm field for a few outputs of differing ionization fractions and
redshifts. The results of this calculation are shown in Figure \ref{fig:rco_lae}. Comparing
the cross spectra of the Ly-$\alpha$ selected galaxies ({\em top panel}, dashed lines) and uniformly 
selected galaxies (solid lines), we
see that the large-scale amplitude of the cross spectra are higher, 
and that the signal turns over on larger scales, for the Ly-$\alpha$ selected galaxy
samples. It is easy to understand these trends qualitatively.
For a galaxy to be visible
in Ly-$\alpha$ it must reside in a sufficiently large H II region -- or more accurately, it needs to reside
along a sufficiently long ionized {\em skewer} -- to avoid complete attenuation owing
to damping wing absorption. 
The largest H II regions form around the most clustered sources, and
so the galaxies detectable in Ly-$\alpha$ are more clustered than uniformly selected 
galaxies of the same host halo mass 
(Furlanetto et al. 2006c, McQuinn et al. 2007a, 2007b, Mesinger \& Furlanetto 2007b, 2007c). This 
enhanced clustering is reflected
in the boosted large scale 21 cm-galaxy cross power spectrum. 
Likewise, the turnover on small scales
is set by the characteristic H II region size around {\em detectable} galaxies, which increases for the
Ly-$\alpha$ selected galaxies: Ly-$\alpha$ galaxies residing in small bubbles are attenuated
out of the sample by damping wing absorption. 
  
This is visualized most clearly in the cross-correlation coefficient between the 21 cm and galaxy
fields (Figure \ref{fig:rco_lae}, {\em bottom panel}). In the uniformly-selected galaxy sample,
the correlation coefficient turns over on progressively larger scales as reionization proceeds. By contrast,
the small scale turnover in the Ly-$\alpha$ selected sample is relatively fixed. At early times when
the bubbles are small, the
turnover in the cross spectrum with the Ly-$\alpha$ selected sample 
{\em largely reflects the damping wing scale}. In order to best characterize bubble growth in the early
and middle stages of reionization, one requires a uniformly-selected galaxy sample, rather than a Ly-$\alpha$
selected sample. Finally, note that the cross correlation coefficient in the Ly-$\alpha$ selected samples
does not quite reach $r=-1$ on large scales. This behavior is enhanced early in reionization, and
presumably results because small bubbles are missed by Ly-$\alpha$ selection, which contribute most
significantly at low ionized fractions. 
In summary, while Ly-$\alpha$ selected samples will be interesting for initial cross 
spectrum detections, 
uniformly-selected samples will be required to best constrain bubble growth during reionization.

\section{Detectability}
\label{sec:detectability}

In this section we calculate the statistical significance
at which future surveys can detect the 21 cm-galaxy cross spectrum and briefly consider the 
resulting insights
into reionization. Here we follow closely the calculations in Furlanetto \& Lidz (2007), simply
extending them to incorporate our simulated cross spectrum signal.
In our calculations we consider a 21 cm survey with the specifications planned for each of the
MWA (Bowman et al. 2006, McQuinn et al. 2006, Mao et al. 2008) and LOFAR, which we review below
(\S \ref{sec:mwa_survey} and \S \ref{sec:lofar}). We consider two basic types of galaxy surveys. First,
we consider a survey similar to the Subaru deep field survey for Ly-$\alpha$ emitters (Kashikawa 2006).
Since the Subaru survey is ongoing, this calculation should illustrate what is achievable in the near
future as the MWA and LOFAR come online. Note that the present Subaru deep field does not overlap
with the planned MWA target fields (M. Morales, private communication, 2008), but our calculations
still serve to illustrate what is possible in the near future. 
Next, we consider a more futuristic galaxy survey. Coupling our futuristic galaxy survey with the
MWA or LOFAR, one can potentially measure the cross power spectrum at several redshifts, probing the
{\em evolution} in the cross spectrum signal, and tracing the growth of H II regions during
reionization as in Figure \ref{fig:cross_v_z}.

\subsection{Statistical Error Estimates}
\label{sec:errorbar}

To begin with, we describe our statistical error estimates, reviewing the formulae for cross spectrum 
error bars for a survey of given specifications, incorporating sample variance, thermal noise in the 21 cm
radio telescope, and shot-noise and redshift errors in the galaxy distribution.
Here we restrict ourselves to the spherically averaged
cross spectrum, since the MWA and LOFAR have limited transverse sensitivity and since very precise
galaxy redshifts will be required to measure the angular dependence of the cross 
spectrum (Furlanetto \& Lidz 2007).

We generally find it convenient to estimate error bars on the cross-correlation 
coefficient, $r(k)$, rather
than on the cross spectrum itself. We desire an estimate of the error bar on $r(k)$ calculated
from spherically averaged auto and cross spectra in a bin of logarithmic width $\epsilon = d\rm{ln}k$.
For notational convenience let us denote the cross-correlation coefficient by
$r(k) = P_{\rm 21, gal}(k)/[P_{\rm 21}(k) P_{\rm gal}(k)]^{1/2} = 
A(k)/[B(k) C(k)]^{1/2}$. Propagating errors, the fractional error on the 
cross-correlation coefficient is:
\beqa
\frac{\sigma_r^2}{r^2}(k) = && \frac{\sigma_A^2}{A^2}(k) + \frac{\sigma_B^2}{4 B^2}(k) +
\frac{\sigma_C^2}{4 C^2}(k) - \frac{\sigma_{AB}^2}{A B}(k) \nonumber \\
&& - \frac{\sigma_{AC}^2}{A C}(k) + \frac{\sigma_{BC}^2}{2 B C}(k).
\label{eq:error_rco}
\eeqa
This expression involves the the cross spectrum variance, the 21 cm and galaxy
power spectrum variances, and the co-variance between the various power spectra, each
calculated for spherically averaged power spectra in shells of logarithmic width
$\epsilon$. The one disadvantage of considering the cross-correlation coefficient
is that the cross spectrum can, under appropriate circumstances, be detected at higher
sensitivity than the 21 cm auto spectrum (Furlanetto \& Lidz 2007). In this case,
the error bar on the cross correlation coefficient, which includes an error term from
the auto spectrum, will be larger than that for the cross spectrum alone. Furthermore,
estimating the cross-correlation coefficient requires an auto spectrum estimate and is hence 
more susceptible to residual foreground contamination than the auto spectrum alone. 

Consider first the power spectrum variance terms for a single $\k$-mode, with line
of sight component $k_\parallel = \mu k$, restricting
ourselves to modes in the upper-half plane. The power spectrum variance expressions
are as follows
(Furlanetto \& Lidz 2007):
\beqa
\sigma_A^2(k,\mu) && = {\rm var}\left[P_{\rm 21, gal} (k,\mu)\right] \nonumber \\&&= \frac{1}{2} 
\left[P_{\rm 21,gal}^2(k,\mu) + \sigma_B(k,\mu) \sigma_C(k,\mu)\right], 
\label{eq:var_cross}
\eeqa

\beqa
\sigma_B^2(k,\mu) && = {\rm var}\left[P_{\rm 21} (k, \mu)\right]  \nonumber \\ 
&& = \left[P_{\rm 21}(k, \mu) + \frac{T^2_{\rm sys}}{T_0^2} \frac{1}{B t_{\rm int}}
\frac{D^2 \Delta D}{n(k_\perp)}\left(\frac{\lambda^2}{A_e}\right)^2\right]^2, \nonumber \\
\label{eq:var21}
\eeqa

\beqa
\sigma_C^2(k, \mu) && =  {\rm var}\left[P_{\rm gal} (k, \mu)\right] \nonumber \\ 
&& = \left[P_{\rm gal}(k, \mu) + n^{-1}_{\rm gal} e^{k^2_\parallel \sigma^2_\chi}\right]^2.
\label{eq:vargal}
\eeqa
The second term in Equation (\ref{eq:var21}) comes from thermal noise in the radio telescope,
the second term in Equation (\ref{eq:vargal}) expresses the shot noise error, while the other terms
in the above equations are sample variance contributions. The thermal noise term 
depends on the system temperature,
$T_{\rm sys}$; the co-moving distance to the center of the survey at redshift $z$, $D(z)$; the
survey depth, $\Delta D$; the observed wavelength, $\lambda$; the effective area of each
antenna tile, $A_e$; the survey bandwidth, $B$; the total observing time, $t_{\rm int}$; and
the distribution of antennas. 
The factor $T_0$ in the denominator of the detector noise term arises because we 
normalize the 21 cm field by $T_0$ so that it is dimensionless -- i.e., we work with the
field $\delta_T/T_0$.
The dependence on antenna configuration is encoded in 
$n(k_\perp)$ which denotes the number density of baselines observing a mode with transverse
wavenumber $k_\perp$ (McQuinn et al. 2006, Bowman et al. 2006, Lidz et al. 2007b). The 
galaxy shot-noise term depends on $n_{\rm gal}$
which is the abundance of galaxies observable in our hypothetical survey, and on the accuracy
of the galaxy redshifts obtained by the survey. The galaxy redshift error is given in co-moving
units by $\sigma_\chi = c \sigma_z/H(z)$. 

We also require expressions for the co-variance between the different power spectra.
These can be computed straightforwardly:
\beqa
\sigma_{AB}^2(k,\mu) && =  {\rm cov}\left[P_{\rm 21, gal} (k,\mu), P_{\rm 21} (k,\mu)\right] \nonumber \\ && = 
 P_{\rm 21, gal}(k,\mu) P_{\rm 21}(k,\mu), \\ 
\sigma_{AC}^2(k,\mu) && =  {\rm cov}\left[P_{\rm 21, gal} (k,\mu), P_{\rm gal} (k, \mu)\right] \nonumber \\ && = 
 P_{\rm 21, gal}(k,\mu) P_{\rm gal}(k), \\ 
\sigma_{BC}^2(k,\mu) && =  {\rm cov}\left[P_{\rm 21} (k,\mu), P_{\rm gal} (k,\mu)\right] \nonumber \\ && =
P_{\rm 21, gal}^2(k,\mu). 
\label{eq:covs}
\eeqa

Finally, we can estimate the error bar on the cross-correlation coefficient formed from our
spherically averaged power spectra. We do this by adding the power spectrum error bars for individual
$\k$-modes from Equations (\ref{eq:var_cross})--(\ref{eq:covs}) in inverse
quadrature, performing a similar calculation for each individual term in 
Equation (\ref{eq:error_rco}).
For example, the variance of the cross-spectrum averaged over 
a spherical shell of logarithmic width $\epsilon = d\rm{ln}k$ is:
\beqa
\frac{1}{\sigma_A^2(k)} = \sum_\mu \frac{\epsilon k^3 V_{\rm survey}}{4 \pi^2}
\frac{\Delta \mu}{\sigma_A^2(k,\mu)}.
\label{eq:var_shell}
\eeqa
The effective survey volume for our radio telescope 
is $V_{\rm survey} = D^2 \Delta D \left(\lambda^2/A_e\right)$. If the galaxy survey has
a lesser volume, $V_{\rm gal}$, then the variance of the binned power spectrum estimated
from this lesser volume (for a mode contained within the lesser survey volume) is larger 
by a factor of $\sim V_{\rm gal}/V_{\rm survey}$.

\subsection{The MWA}
\label{sec:mwa_survey}

With these expressions in hand let us briefly describe the specifications 
we assume for our 21 cm and galaxy surveys. The MWA will have a large field of view, 
spanning $\sim 800$ deg$^2$ on the sky, 
and consisting of $500$ antenna tiles each
with an effective area of $A_e = 14 m^2$ at $z = 8$ (Bowman et al. 2006). Each antenna tile is $4 m$ wide,
and we follow Bowman et al. (2006), McQuinn et al. (2006) in assuming that the antennas are packed
as closely as possible within a compact core, with the distribution subsequently falling off as $r^{-2}$ 
in order to capture large baselines, out to a maximum baseline of $1.5$ km. 
Lidz et al. (2007b) argued that a compact antenna configuration, with all of the MWA's 
antennas packed as close as possible, is a superior configuration for 21 cm auto spectrum
measurements. This configuration is less good for the cross spectrum: given a galaxy
survey with photometric redshifts, one needs to balance the MWA's high
line-of-sight sensitivity, yet poor transverse sensitivity, with the galaxy survey's
high transverse sensitivity, yet poor line-of-sight sensitivity owing to redshift 
uncertainties.  
  
We assume that the
system temperature is set by the sky temperature, which we take to be 
$T_{\rm sys} = 280 (1+z/7.5)^{2.3}$ K, following Wyithe \& Morales (2007). We consider
a bandwidth of $B = 6$ Mhz observing for a total time of $t_{\rm int} = 1000$ hrs. The bandwidth
is chosen to be small enough to ensure that the signal evolves minimally over the corresponding 
redshift interval (McQuinn et al. 2006).

\subsection{LOFAR}
\label{sec:lofar}

LOFAR and the MWA are expected to have comparable sensitivity for 
detecting the 21 cm auto spectrum (McQuinn et al. 2006, Mao et al. 2008).\footnote{We recently
learned that budget setbacks are forcing LOFAR to reduce its collecting area. We are unaware of
the details of the reduction, but this will reduce LOFAR's sensitivity compared to our
estimates here.} 
LOFAR will observe
a smaller field of view than the MWA (by a factor of $\gtrsim 10$), but its larger collecting area 
compensates for its reduced sky coverage. 
The larger field of view of the MWA is, however, wasted when 
cross-correlating with
a galaxy survey that covers a much smaller patch on the sky. We anticipate
then that LOFAR should at least initially 
provide a {\em more sensitive} detection
of the 21 cm-galaxy cross spectrum than the MWA (Furlanetto \& Lidz 2007).

The precise collecting area and antenna configuration for LOFAR are 
still evolving, but we follow the simple model of McQuinn et al. (2006) as a 
plausible estimate. LOFAR
will consist of $32$ large antenna stations within $1$ km, with minimum
baselines of $100$ m. Each LOFAR station simultaneously 
observes $4$ separate regions on the sky.
We assume that LOFAR's antenna stations are closely packed in
a compact core, before tapering off in an $r^{-2}$ configuration out to
a maximum radius of $1$ km. The effective area of each 
antenna is $656$ m$^2$ at $z=8$, and
we linearly interpolate between the values in McQuinn et al. (2006)
(their Table 1), to find the
collecting area at other redshifts. As for the MWA, we consider $1,000$ hrs. of LOFAR
observations over a bandwidth of $B=6$ Mhz.

\subsection{Subaru-like Survey}
\label{sec:survey_subaru}

We first consider the detectability of the 21 cm-galaxy cross spectrum obtainable by 
combining the MWA and LOFAR with the Subaru deep field survey, and plausible extensions. The 
existing
Subaru deep field survey has a $0.25$ deg$^2$ field of view and locates Ly-$\alpha$
emitters near $z=6.6$ to a depth of $130$ \AA. 
The existing spectroscopically-confirmed Subaru deep
field sample at redshift $z=6.6$ consists of $36$ emitters (Kashikawa et al. 2006). 
The number density of spectroscopically-confirmed emitters corresponds to 
$n_{\rm gal} = 1.6 \times 10^{-4}$ Mpc$^{-3}$. 
An extension to the Subaru deep field, the Subaru/XMM-Newton Deep Survey is already underway,
and promises to increase the observed $z=6.6$ field of view by a factor of $\sim 4$, reaching
a survey area of $A_{\rm survey} \sim 1$ deg$^2$ by the end of the year (Ouchi 2005). 

Given the rapid progress in area surveyed, we examine how the 
detectability of the cross spectrum scales
with increasing field-of-view, at fixed depth and galaxy number density. In practice,
we calculate the cross spectrum $S/N$ for a galaxy survey that covers
the full field-of-view of the MWA ($\sim 800$ deg$^2$ at this redshift), and scale the
signal-to-noise (squared) in each $k$-bin downwards by the 
ratio of the galaxy survey volume to
that of the MWA (see
Equation \ref{eq:var_shell}). We perform a similar calculation for LOFAR. 
For each model cross-spectrum, our $S/N$ estimates assume
a galaxy number density of 
$n_{\rm gal} = 1.6 \times 10^{-4}$ Mpc$^{-3}$, and redshift errors for the spectroscopically
confirmed galaxies of $\sigma_z = 0.01$. 
The assumed redshift error corresponds to a velocity of several hundred km/s, motivated
by the typical velocity offsets for Ly-$\alpha$ lines observed by Shapley et al. (2003) 
in Lyman break galaxies at $z \sim 3$.
The total $S/N$ is determined by summing the signal squared divided by our variance estimate (Equations
\ref{eq:var_cross} and \ref{eq:var_shell}) over all detected $k$-bins.

\begin{figure}
\bc
\includegraphics[width=9.2cm]{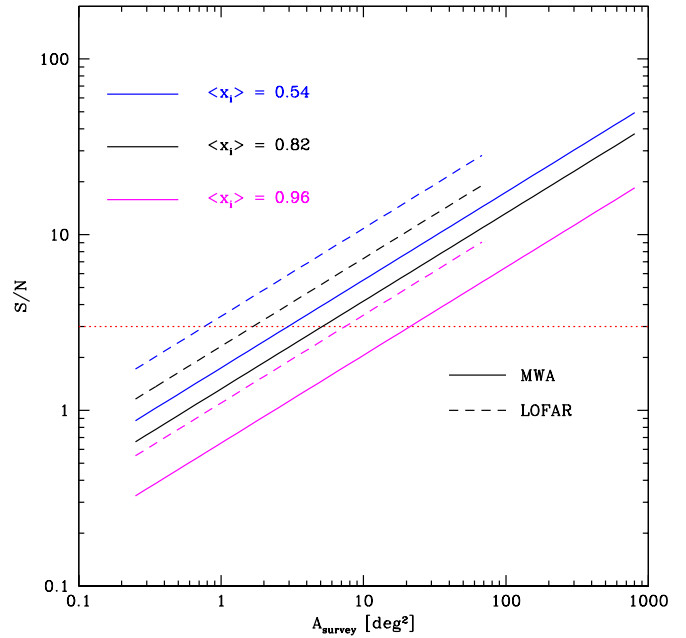}
\caption{Signal to noise for cross spectrum detection. The signal to noise
at which Subaru-like surveys, coupled with the MWA (solid lines) and LOFAR (dashed lines), 
can detect the 21 cm-galaxy cross spectrum as 
a function of survey area at $z = 6.6$. The different curves indicate different models
for the ionization fraction. Each curve extends from the current Subaru
area ($\sim 0.25$ deg$^2$) to the full field of view of MWA ($\sim 800$ deg$^2$) or
LOFAR ($\sim 70$ deg$^2$).
The red dotted line indicates a $3-\sigma$ detection of the cross spectrum.}
\label{fig:ston}
\ec
\end{figure}

In Figure \ref{fig:ston} we show the detectability of the cross spectrum for 
a few different models over a range of survey areas. In each model we adopt a plausible 
minimum detectable
galaxy mass of $M_{\rm g, min} = 10^{10} M_\odot$, fixing the galactic duty cycle
to match the observed Subaru deep field abundance of
$n_{\rm gal} = 1.6 \times 10^{-4}$ Mpc$^{-3}$. The corresponding duty cycle in our models is around
$\sim 1\%$.
Given that we currently have few
direct observational constraints on the filling factor of H II regions near
$z = 6.6$, we consider models in which the ionized fraction is $\avg{x_i} = 0.54, 0.82$,
and $0.96$ at this redshift. Strictly speaking, we should use our Ly-$\alpha$ selected
cross spectrum models here, but at these ionization fractions we expect this to boost 
our $S/N$ only
slightly (Figure \ref{fig:rco_lae}). For simplicity, we conservatively ignore the 
clustering-boost from Ly-$\alpha$ selection here.
The amplitude of the cross spectrum
is largest amongst these models at $\avg{x_i} = 0.54$, and is substantially smaller
by $\avg{x_i} = 0.96$ (see Figure \ref{fig:cross_v_z}), and so the more neutral
models will be easier to detect. 

The results shown in Figure \ref{fig:ston} illustrate
that cross-correlating the MWA with a galaxy survey of size comparable to the
present Subaru deep field survey will not allow a significant cross spectrum detection 
($S/N \lesssim 1-\sigma$), even if the IGM is significantly neutral at $z=6.6$. However,
extensions to the Subaru deep field that  
cover a larger area on the sky should yield significant cross spectrum
detections. For example, extending the present sky coverage by a factor of $\sim 10-15$ to
$3$ deg$^2$ should provide a $\gtrsim 2-3 \sigma$ cross spectrum detection in our 
$\avg{x_i} = 0.54$ and $\avg{x_i} = 0.82$ models, but only 
a $\sim 1-\sigma$ detection in our $\avg{x_i} = 0.96$ model. 
As anticipated, cross-correlating with LOFAR can improve the $S/N$ by a factor of a few.
Cross-correlating LOFAR with a galaxy survey of only $1-2$ deg$^2$ should allow a $3-\sigma$
cross spectrum detection in our $\avg{x_i} = 0.54$ and $\avg{x_i} = 0.82$ models. 
As mentioned earlier, the Subaru survey should reach this sky coverage soon, making a cross
spectrum detection feasible in the next few years if the IGM is partly neutral around $z \sim 6.6$. 
More ambitious surveys covering
the entire MWA sky $\sim 800$ deg$^2$ would clearly move beyond mere detections --
the detection $S/N$ for such surveys is at the tens of sigma level 
(see Figure \ref{fig:ston}) -- and provide valuable constraints on reionization models. 

\subsection{Futuristic Survey}
\label{sec:survey_future}

Since more futuristic surveys will go beyond mere detections, we proceed to consider
the constraining power of a large field-of-view galaxy survey -- cross-correlated with
the MWA -- in more 
detail. 
Futuristic surveys will allow one to probe small scales, capture the turnover in the
cross-correlation coefficient and hence constrain bubble growth during reionization. 
We calculate
the expected error bar on the cross correlation coefficient as a function of wavenumber 
for a galaxy survey spanning the full MWA field of view, and consider the ability
of this survey to constrain reionization models. Here we assume that the galaxy survey
can detect fainter galaxies, reaching a galactic abundance $100$ times larger than in the previous
section, with the same redshift accuracy of $\sigma_z = 0.01$. We consider a redshift
of $z = 7.3$.

\begin{figure}
\bc
\includegraphics[width=9.2cm]{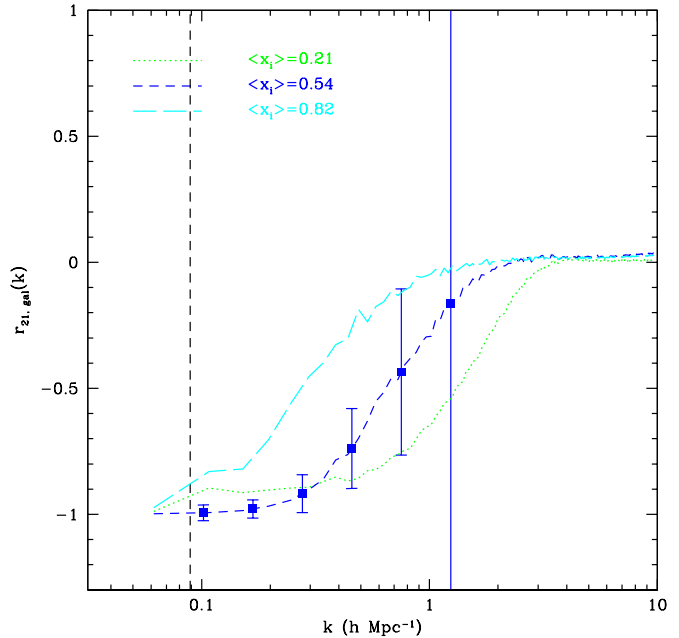}
\caption{Error estimate for the 21 cm-galaxy cross correlation coefficient. Here we
consider a futuristic galaxy survey covering the entire MWA field of view, cross-correlated
with the MWA.
The blue points show the mean signal and error estimates for our hypothetical
21 cm/galaxy survey when $\avg{x_i} = 0.54$. The other curves show
the cross correlation coefficient when $\avg{x_i} = 0.21$ and $0.82$ 
respectively. Our hypothetical survey should help constrain the
volume-weighted ionization fraction. The vertical black dashed line shows
the wavenumber corresponding to the survey depth, below which foreground
contamination will prohibit extracting the signal.
}
\label{fig:rco_detect}
\ec
\end{figure}

Using again the models of \S \ref{sec:evol_sig} as input, we 
estimate the statistical sensitivity of our futuristic galaxy survey.
The results of our sensitivity calculation are shown in 
Figure \ref{fig:rco_detect}, for spherical bins of logarithmic width $\epsilon = 0.5$. 
Here we plot the simulated signal when the IGM
is $\sim 50\%$ ionized along with a statistical error estimate for our 
hypothetical survey. For contrast, we additionally show theoretical model
curves when the IGM is each of $\sim 20\%$ and $\sim 80\%$ ionized. 

The curves and errorbars in Figure \ref{fig:rco_detect} show that the statistical precision of
our hypothetical survey is high enough to distinguish between the different stages of reionization
shown in the figure over about a decade in scale. On large scales, the measurement is limited
by foreground removal while on small scales 21 cm detector noise and galaxy redshift errors limit
the statistical precision of the measurements (Furlanetto \& Lidz 2007).
Although the 21 cm-galaxy cross power spectrum signal is much less susceptible than the
21 cm auto power spectrum to foreground contamination, free-free and synchrotron emission from
the high redshift galaxies in our survey still contaminate the 21 cm-galaxy cross power spectrum
somewhat (Furlanetto \& Lidz 2007). 
This prohibits measuring modes with lines of sight wave numbers
$k_\parallel < 2 \pi/\Delta D$, where $\Delta D$ is the depth of the survey. The discreteness of
the survey means that the only modes in our survey that satisfy this requirement 
have $k_\parallel = 0$, hence
all modes with $k \leq k_\parallel$ will be removed in the foreground cleaning process.
The black dashed line indicates the wavenumber corresponding to the survey depth in our 
hypothetical
survey. We should remind the reader here of one trade-off involved with considering the cross-correlation
coefficient rather than the cross spectrum alone.
The cross-correlation coefficient is a more convenient quantity than the cross spectrum
for visualizing the small-scale turn-over (see Figure \ref{fig:rco_detect}), but it is 
less desirable in that it includes the
auto spectrum, which is more susceptible to foreground contamination.

The sensitivity estimates shown in Figure \ref{fig:rco_detect} are encouraging, and suggest that
future 21 cm-galaxy surveys may help constrain the filling factor 
and size distribution of H II regions during reionization. Comparing our error estimates
with the results of Figure \ref{fig:rco_galmass} suggests that futuristic surveys might also -- by
measuring the cross spectrum in different galaxy luminosity bins -- weakly
constrain the dependence of bubble size on host halo mass. Note also that the thermal noise term in the
21 cm variance (see Equation \ref{eq:var21}) still contributes significantly for most $k$-bins shown here,
and so futuristic 21 cm surveys with more antennas and larger collecting areas than the MWA can further 
improve cross spectrum sensitivity. In particular, a future FFT telescope (Tegmark \& Zaldarriaga 2008) should
boost the sensitivity compared to our estimates here (see Mao et al. 2008 for estimates of the auto spectrum
sensitivity with an FFT telescope).

\section{Conclusions} \label{sec:conclusions}

In this paper, we considered the scientific return of future 21 cm-galaxy cross
power spectrum measurements. A strong cross spectrum measurement ultimately
requires detecting a sizable number of high redshift galaxies over a large field of
view, which presents a significant observational challenge. Nonetheless, we showed
that a detection of the cross spectrum may be achieved in the near future by combining LOFAR and 
the Subaru survey for LAEs at $z \sim 6.6$. We estimate that a $\sim 3-\sigma$ detection is
feasible, provided the IGM is $\gtrsim 20\%$ neutral at this redshift, and that the Subaru
survey's sky coverage is extended
from $0.25$ deg$^2$ to $\sim 2$ deg$^2$. This detection would already be quite valuable,
as it would help confirm that the detected 21 cm signal comes from the high redshift IGM, and
not from foreground contamination, which should mostly be uncorrelated with high redshift
galaxies (Furlanetto \& Lidz 2007).

Futuristic galaxy surveys covering $100$s of square degrees on the sky, can be combined
with the MWA, LOFAR, and other 21 cm surveys, to move beyond a mere detection of the 
cross spectrum signal and map out its detailed scale dependence. 
The galaxy surveys required for these measurements are clearly very challenging, but
rapid progress is being made in this direction as deep, widefield surveys  
are being designed to study baryonic acoustic oscillations and/or weak-lensing at high redshift
(e.g., ADEPT, HETDEX\footnote{http://www.as.utexas.edu/hetdex/}, CIP, and others). Another
option is to sparsely sample the MWA or LOFAR fields, in order to capture the large-scale modes
(Furlanetto \& Lidz 2007).
We have shown that the 21 cm-galaxy cross spectrum is a relatively direct tracer of bubble
growth during reionization. Measuring the turnover scale as a function of galaxy luminosity
constrains the luminosity dependence of the characteristic bubble size. This information is
difficult, or impossible, to obtain with the 21 cm auto spectrum alone.
In order to extract the most
information out of the cross spectrum, it should be combined with measurements of the
galaxy auto spectrum and luminosity function, which will help to constrain the galaxy
luminosity-halo mass correlation. 

A further interesting feature of the simulated
signal is that the cross-correlation changes sign on large scales near the beginning
of reionization (Figure \ref{fig:cross_v_z}). At this early phase of reionization, our 
results may, however, be modified by spin temperature fluctuations, which we presently neglect. 
Future work should incorporate these fluctuations. If our signature holds up, the change in
sign of the cross correlation would provide a very interesting observational 
indicator of the earliest phases of reionization. Finally, we found that the 21 cm-galaxy
cross power spectrum might provide an interesting observational signature of scenarios where 
ionizing
photons fail to escape from low mass halos. Provided galaxies in these
low mass halos are detectable longward of the ionization edge, we expect the cross spectrum
to change sign and turn positive on small scales. 
Generally speaking, the 21 cm-galaxy cross 
spectrum
is a more direct tracer of the impact of galaxies on the surrounding IGM than the 21 cm auto 
spectrum.
As such, it can potentially provide a wealth of information about the EoR and early structure 
formation.

\section*{Acknowledgments}
We thank Mark Dijkstra and Miguel Morales for helpful discussions. We thank Suvendra Dutta
for useful conversations and for his collaboration in related work. 
Support was provided, in part, by the David and Lucile Packard Foundation, the
Alfred P. Sloan Foundation, and grants AST-0506556 and NNG05GJ40G. OZ 
acknowledges additional support by a Berkeley Center for
Cosmological Physics (BCCP) Fellowship.




\begin{thebibliography}{}

\bibitem[]{Abel02} Abel, T., \& Wandelt, B.~D. 2002, MNRAS, 330, 53

\bibitem[]{Barkana01} Barkana, R., \& Loeb, A. 2001, Phys. Rept., 349, 125

\bibitem[]{Barkana02} Barkana, R., \& Loeb, A. 2002, ApJ, 578, 1 

\bibitem[]{Bouwens08} Bouwens, R.~J., Illingworth, G.~D., Franx, M., \& Ford, H. 2008,
ApJ submitted, arXiv:0803.0548

\bibitem[]{Bowman06} Bowman, J.~D., Morales, M.~F., \& Hewitt, J.~N. 2006, ApJ, 638, 20

\bibitem[]{Chang08} Chang, T.-C., Pen, U.-L., Peterson, J.~B., \& McDonald, P. 2008, Phys. Rev. L, in press, arXiv:0709.3672 

\bibitem[]{Ciardi03} Ciardi, B., \& Madau, P. 2003, ApJ, 596, 1

\bibitem[]{Crocce06} Crocce, M., Pueblas, S., Scoccimarro, R. 2006, MNRAS, 373, 369
05

\bibitem[]{Dijkstra04} Dijkstra, M., Haiman, Z., Rees, M.~J., \& Weinberg, D.~H. 2004, ApJ, 601, 666
 
\bibitem[]{Dijkstra07} Dijkstra, M., Lidz, A., \& Wyithe, J.~S.~B. 2007, MNRAS, 377, 1175

\bibitem[]{Furlanetto04} Furlanetto, S.~R., Zaldarriaga, M., \& Hernquist, L. 2004, ApJ, 613, 1

\bibitem[]{Furlanetto05} Furlanetto, S.~R., \& Oh, S.~P., 2005, MNRAS, 363, 1031

\bibitem[]{Furlanetto06a} Furlanetto, S.~R., Oh, S.~P., \& Briggs, F. 2006a, Phys. Reports, 433, 181

\bibitem[]{Furlanetto06b} Furlanetto, S.~R., McQuinn, M., \& Hernquist, L. 2006b, MNRAS, 365, 115

\bibitem[]{Furlanetto06c} Furlanetto, S.~R., Zaldarriaga, M., \& Hernquist, L. 2006c, MNRAS, 365, 1012

\bibitem[]{Furlanetto07} Furlanetto, S.~R., \& Lidz, A. 2007, ApJ, 660, 1030

\bibitem[]{Gnedin08} Gnedin, N.~Y., Kravtsov, A.~V., \& Chen, H-W 2008, ApJ, 672, 765

\bibitem[]{Haiman01} Haiman, Z., Abel, T., \& Madau, P. 2001, ApJ, 551, 599

\bibitem[]{Kashikawa06} Kashikawa, N., et al. 2006, ApJ, 648, 7

\bibitem[]{Komatsu08} Komatsu, E., et al. 2008, ApJS submitted, arXiv:0803.0547

\bibitem[]{Lidz07a} Lidz, A., Zahn, O., McQuinn, M., Zaldarriaga, M., Dutta, S., \& Hernquist, L.
2007a, ApJ, 659, 865

\bibitem[]{Lidz07b} Lidz, A., Zahn, O., McQuinn, M., Zaldarriaga, M., \& Hernquist, L.
2007b, ApJ in press, arXiv:0711.4373 

\bibitem[]{Madau97} Madau, P., Meiksin, A., \& Rees, M.~J. 1997, ApJ, 475, 429

\bibitem[]{Mao08} Mao, Y., Tegmark, M., McQuinn, M., Zaldarriaga, M., \& Zahn, O. 2008, PRD submitted,
arXiv:0802.1710

\bibitem[]{McQuinn06} McQuinn, M., Zahn, O., Zaldarriaga, M., Hernquist, L., \& Furlanetto, S.~R.
2006, ApJ, 653, 815

\bibitem[]{McQuinn07a} McQuinn, M., Lidz, A., Zahn, O., Dutta, S., Hernquist, L., \& Zaldarriaga, M.
2007a, MNRAS, 377, 1043

\bibitem[]{McQuinn07b} McQuinn, M., Hernquist, L., Zaldarriaga, M., \& Dutta, S. 2007b, MNRAS, 381, 75

\bibitem[]{Mesinger07a} Mesinger, A., \& Furlanetto, S.~R., 2007a, ApJ, 669, 663

\bibitem[]{Mesinger07b} Mesinger, A., \& Furlanetto, S.~R., 2007b, MNRAS in press, arXiv:0708.0006

\bibitem[]{Mesinger07c} Mesinger, A., \& Furlanetto, S.~R., 2007c, MNRAS submitted, arXiv:0710.0371

\bibitem[]{Miralda98} Miralda-Escud\'e, J. 1998, ApJ, 501, 15

\bibitem[]{Miralda00} Miralda-Escud\'e, J., Haehnelt, M., \& Rees, M.~J. 2000, ApJ, 530, 1

\bibitem[]{Navarro97} Navarro, J.~F., \& Steinmetz, M. 1997, ApJ, 478, 13

\bibitem[]{Oh03} Oh, S.~P., \& Haiman, Z. 2003, MNRAS, 346, 456

\bibitem[]{Ouchi05} Ouchi., M., et al., 2005, ApJ, 620, L1

\bibitem[]{Partridge67} Partridge, R.~B., \& Peebles, P.~J.~E. 1967, ApJ, 147, 868

\bibitem[]{Pritchard07} Pritchard, J.~R., \& Furlanetto, S.~R. 2007, MNRAS, 376, 1680

\bibitem[]{Pritchard08} Pritchard, J.~R., \& Loeb, A. 2008, Phys. Rev. D, submitted, 2008,
arXiv:0802.2102

\bibitem[]{Rhoads04} Rhoads, J.~E., Xu, C., Dawson, S., Dey, A., Malhotra, S., Wang, J.-X., Jannuzi, B.~T., 
Spinrad, H., \& Stern, D. 2004, ApJ, 611, 59

\bibitem[]{Scoccimarro01} Scoccimarro, R., Sheth, R.~K., Hui, L., \& Jain, B. 2001, ApJ, 546, 20

\bibitem[]{Scott90} Scott, D., \& Rees 1990, MNRAS, 247, 510

\bibitem[]{Shapiro04} Shapiro, P.~R., Iliev, I.~T., \& Raga, A.~C. 2004, MNRAS, 348, 753

\bibitem[]{Shapley03} Shapley, A.~E., Steidel, C.~C., Pettini, M., \& Adelberger, K.~L.
2003, ApJ, 588, 65

\bibitem[]{sokasian01} Sokasian, A., Abel, T., \& Hernquist, L. 2001, NewA, 6,
359
 
\bibitem[]{sokasian03} Sokasian, A., Abel, T., Hernquist, L. \&
Springel, V. 2003, MNRAS, 344, 607

\bibitem[]{Spergel07} Spergel, D.~N., et al., 2007, ApJS, 170, 377
 
\bibitem[]{springel05} Springel, V. 2005, MNRAS, 364, 1105

\bibitem[]{Springel03} Springel, V., \& Hernquist, L. 2003, MNRAS, 339, 312

\bibitem[]{Stark07} Stark, D.~P., Ellis, R.~S., Richard, J., Kneib, J.-P., Smith, G.~P., \&
Santos, M.~R. 2007, ApJ, 663, 10

\bibitem[]{Tegmark08} Tegmark, M., \& Zaldarriaga, M. 2008, Phys. Rev. D submitted, arXiv:0805.4414
 
\bibitem[]{Thoul96} Thoul, A.~A., \& Weinberg, D.~H. 1996, ApJ, 465, 608

\bibitem[]{Wyithe07} Wyithe, J.~S.~B., \& Loeb, A. 2007, MNRAS, 375, 1034

\bibitem[]{Wyithem07} Wyithe, J.~S.~B., \& Morales, M. 2007, MNRAS submitted, astro-ph/0703070

\bibitem[]{Wyithe08} Wyithe, J.~S.~B., \& Loeb, A. 2008, MNRAS submitted, arXiv:0708.3392

\bibitem[]{zahn07} Zahn, O., Lidz, A., McQuinn, M., Dutta, S., Hernquist, L., Zaldarriaga, M., \& Furlanetto, S.~R. 2007, ApJ, 654, 12

\bibitem[]{Zaldarriaga04} Zaldarriaga, M., Furlanetto, S.~R., \& Hernquist, L. 2004, 
ApJ, 608, 622

\end{thebibliography}
\end{document}